\documentclass[%
 reprint,
superscriptaddress,
 amsmath,amssymb,
 aps,
prb,
floatfix,
]{revtex4-1}

\pdfoutput=1
\usepackage{color}
\usepackage{graphicx}
\usepackage{placeins}
\usepackage{dcolumn}
\usepackage{bm}
\usepackage{hyperref}

\begin{document}

\title{Exact diagonalization study of the Hubbard-parametrized four-spin ring exchange model on a square lattice}

\author{C.~B.~Larsen}
\affiliation{Nanoscience Center, Niels Bohr Institute, University of Copenhagen, DK-2100 Copenhagen, Denmark}
\affiliation{School of Metallurgy and Materials, University of Birmingham, Edgbaston, Birmingham B15 2TT, United Kingdom}

\author{A.~T.~R\o mer}%
\affiliation{Nanoscience Center, Niels Bohr Institute, University of Copenhagen, DK-2100 Copenhagen, Denmark}
\affiliation{Institut Laue-Langevin, 
71 avenue des Martyrs CS 20156, 38042 Grenoble Cedex 9, France}

\author{S.~Janas}%
\affiliation{Nanoscience Center, Niels Bohr Institute, University of Copenhagen, DK-2100 Copenhagen, Denmark}

\author{F.~Treue}%
\affiliation{Nanoscience Center, Niels Bohr Institute, University of Copenhagen, DK-2100 Copenhagen, Denmark}

\author{B.~M\o nsted}%
\affiliation{Nanoscience Center, Niels Bohr Institute, University of Copenhagen, DK-2100 Copenhagen, Denmark}
\affiliation{Department of Applied Mathematics and Computer Science, Technical University of Denmark, DK-2800 Lyngby, Denmark}

\author{N.~E.~Shaik} 
\affiliation{Laboratory of Quantum Magnetism, Institute of Physics, EPFL, 1015 Lausanne, Switzerland}
\author{H.~M.~R\o nnow}%
\affiliation{Laboratory of Quantum Magnetism, Institute of Physics, EPFL, 1015 Lausanne, Switzerland}
\affiliation{Nanoscience Center, Niels Bohr Institute, University of Copenhagen, DK-2100 Copenhagen, Denmark}

\author{K.~Lefmann}%
\affiliation{Nanoscience Center, Niels Bohr Institute, University of Copenhagen, DK-2100 Copenhagen, Denmark}

\date{\today}

\begin{abstract}
We have used exact numerical diagonalization to study the excitation spectrum and the dynamic spin correlations in the $s=1/2$ next-next-nearest neighbor Heisenberg antiferromagnet on the square lattice, with additional 4-spin ring exchange from higher order terms in the Hubbard expansion. We have varied the ratio between Hubbard model parameters, $t/U$, to obtain different relative strengths of the exchange parameters, while keeping electrons localized. The Hubbard model parameters have been parametrized via an effective ring exchange coupling, $J_r$, which have been varied between 0~$J$ and 1.5~$J$. We find that ring exchange induces a quantum phase transition from the $(\pi, \pi)$ ordered Ne\`el state to a $(\pi/2, \pi/2)$ ordered state. This quantum critical point is reduced by quantum fluctuations from its mean field value of $J_r/J = 2$ to a value of $\sim 1.1$. At the quantum critical point, the dynamical correlation function shows a pseudo-continuum at $q$-values between the two competing ordering vectors.
\end{abstract}
\maketitle

\section{Introduction}
In the cuprate perovskite materials, magnetic fluctuations constitute a main candidate for the glue giving the binding of the Cooper pairs that lead to superconductivity. For this reason, the magnetic properties of these cuprates are under intense investigation\cite{scalapino1999, birgeneau2006}. The cuprate parent compounds are  antiferromagnetically ordered Mott insulators and insights into their magnetic structure and dynamics provides the groundwork for understanding of the magnetic properties of the cuprate superconductors.

To describe the behaviour of a magnetic insulator, one often applies the Heisenberg Hamiltonian
\begin{equation}
 H = - \sum_{i, \tau} J_{\tau} \hat{{\bf S}}_i \cdot \hat{{\bf S}}_{i+\tau},
\end{equation}
where $\hat{{\bf S}}_i$ denotes the spin on lattice site $i$, and $J_\tau$ is the interaction with a neighbor spin at position $i+\tau$. This provides a good model of Mott-insulating systems where electron mobility is prevented due to strong electron-electron repulsion.

In the opposite limit, where electron hopping becomes of the same order as the Coulomb repulsion, the electronic system is usually described by the Hubbard model, which has proven very successful in describing the $d$-wave superconductivity of the hole-doped cuprate system\cite{maier2000}. 

In the intermediate regime, Coulomb repulsion remains the largest energy, but there is an increase in electron mobility which can be expressed in terms of higher order exchange interactions. This leads to the formulation of the so-called Ring Exchange Model (REM) where a plaquette of four spins is involved in the effective interaction Hamiltonian. The REM thus provides an intermediate step between the two limiting cases of the Heisenberg Hamiltonian and the Hubbard Hamiltonian. La$_2$CuO$_4$, the parent compound of the cuprates La$_{2-x}$Ba$_x$CuO$_4$ and La$_{2-x}$Sr$_x$CuO$_4$, falls into this category\cite{coldea2001, katanin2002, katanin2003, toader05, goff2007}. It is essentially two-dimensional, with a quantum spin $s=1/2$ on each site placed in a square geometry, where quantum effects are expected to play a dominant role.

Previous theoretical studies of the REM include the mean field and spin wave studies by Ref.~\onlinecite{chubukov1992}, where it is found that a substantial large ring exchange ($J_r\geq 2$, see eq.~(\ref{eq:H4})) would drive the system out of the N\'eel state and into a state where the two N\'eel sublattices are canted with respect to one another.
This study was continued by Ref.~\onlinecite{majumdar12}, which, using second-order spin-wave theory up to second nearest neighbor, find that the destabilization of the N\'eel order begins already at values of the ring exchange coupling constant around $J_r \sim 1$. A similar effect has also been observed in exact diagonalization (ED) studies\cite{roger89, chubukov1992}, but these studies have been limited to rather small system sizes of up to 16 to 18 spins. Ref.~\onlinecite{misumi14} uses ED of systems up to $N=32$, but uses an unconventional functional form for the ring exchange. In addition, none of the aforementioned studies focus strictly on all necessary interactions  (which we list in Eq.\ (\ref{eq:H4}) shown below), but rather construct a many-parameter model, where each interaction can vary freely. In this respect, the earlier studies do not reflect the nature of a true Hubbard-parametrized ring exchange model. On the other hand, inelastic neutron scattering studies of La$_2$CuO$_4$ have been able to reconcile experimental spin wave dispersions with Hubbard-parametrized linear spin wave dispersions\cite{coldea2001}. This motivates a more systematic numerical study of the Hubbard-parametrized REM.

Exact diagonalization of a finite sized spin Hamiltonian provides all-encompassing information on the quantum state with no symmetry bias. Full knowledge of the quantum ground state makes it possible to calculate the dynamical structure factor, which allows direct comparison between numerical and experimental results\cite{lefmann, ronnow01, stone03, luscher}. Extracting similar information from other approaches such as mean-field studies or quantum Monte Carlo methods is more obscure, because it requires manual breaking of the spin-rotational symmetry and additional assumptions about the line shapes of the excitation spectra\cite{sandvik}. A drawback of the ED method is the excessive computational cost for large system sizes due to the $(2s+1)^{N}$ scaling of the number of elements in the Hamiltonian matrix, with $N$ being the number of spins. Exact diagonalization studies have been reported of 2D systems of $s=1/2$ with up to 64 spins, though these studies have been restricted to quantum states with finite magnetizations\cite{luscher}. Finite magnetization states inhabit smaller Hilbert spaces than zero magnetization states because of lower degrees of freedom. Recently, we performed ED on a $N = 36$ spin 1/2 system in the zero magnetization subspace \cite{ourotherpaper}. The current size record for a 2D system in the zero magnetization subspace is for an N=48 Kagom{\'e} system \cite{lauchli2016}, and for a 1D chain an N=50 system has been benchmarked \cite{weisse2013}. Though the available system sizes are smaller for the zero magnetization subspace, the study of these states is important because the true ground state of the ring exchange model must reside in this subspace due to the strong antiferromagnetic coupling. \\

We here present an ED study of the dispersion and structure factor of the antiferromagnet Hubbard-parametrized ring exchange model in addition to a linear spin wave (LSW) calculation. The present study is dedicated to the determination of quantum effects on the magnetism due to prominent virtual electron hopping on the square lattice. In particular, we will determine the dynamical structure factor, $S(\textbf{q},\omega)$ for different values of the ring exchange strength. Our results are compared to the magnetic excitation spectrum and the corresponding spectral weights deduced from inelastic neutron scattering on La$_2$CuO$_4$.
 
\subsection{Relation between the Hubbard model and the spin-1/2 four-spin ring exchange model}
\begin{figure}
	\centering
    \includegraphics[clip=true,width=6.02cm]{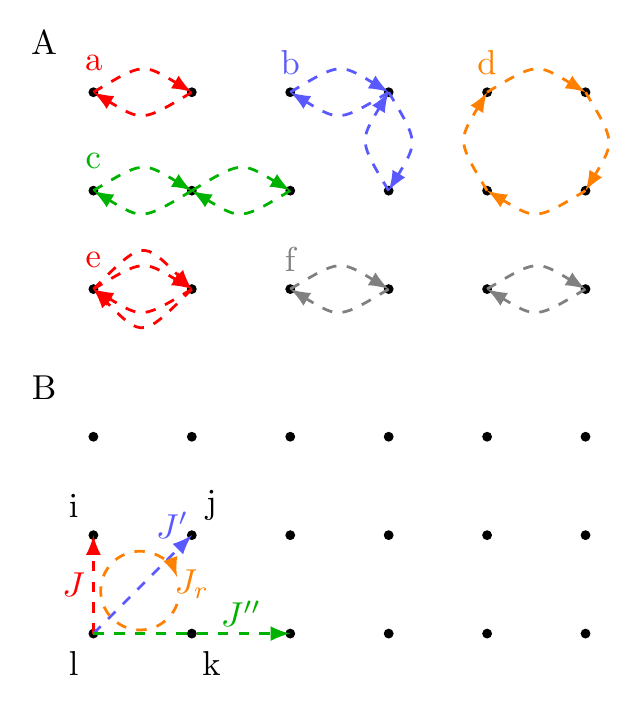}
	\caption{\textbf{A}: Exchange processes behind the parameterization of the ring exchange model. The number of lines of each process corresponds to the number of virtual jumps a given electron was allowed to take during the expansion of the one-band Hubbard model, Eq. (\ref{eq:Hubbard}). Exchange processes a), b), c) and e) give rise to effective Heisenberg couplings between nearest, next-nearest and next-next-nearest neighbors. Process d) is the ring exchange term that couples four spins in a circle, while f) gives two unconnected loops, which cancel. \textbf{B}: Illustration of the effective exchange couplings.}
	\label{fig:ringex}
\end{figure}

The Heisenberg Hamiltonian is derived from a second-order expansion of virtual electron hopping in the Hubbard model
	\begin{equation}
		H = - t\sum_{i,\tau,\sigma} \hat{c}_{i,\sigma}^\dagger \hat{c}_{i+\tau,\sigma} -\mu \sum_i  n_{i,\sigma}+ U \sum_i n_{i,\uparrow} n_{i,\downarrow} \label{eq:Hubbard}
	\end{equation}
	where $\hat{c}_{i,\sigma}^\dagger$ and $\hat{c}_{i, \sigma}$ are the fermionic creation and annihilation operators, $n_{i,\uparrow}$ and $n_{i,\downarrow}$ are the counting operators,  $t$ is the nearest-neighbor hopping matrix element, $\mu$ is the chemical potential and $U$ is the Coulomb repulsion. 
The Heisenberg model is obtained in the limit of vanishing electron mobility, $U \gg t$, in the half-filled system with $\mu=0$.
When the hopping $t$ increases, it becomes relevant to include higher order terms in the expansion and thereby 
transform the Hubbard model to an effective spin Hamiltonian with ring exchange terms. We expand to fourth order in $t$ and project to a subspace with no double occupancies \cite{macdonald, piazza}:
	\begin{eqnarray}
			\hat{H}^{(4)} & = & J \sum_{\langle i,j \rangle}  \hat{\textbf{S}}_i \cdot \hat{\textbf{S}}_j + J' \sum_{\langle i,j'\rangle} \hat{\textbf{S}}_i \cdot \hat{\textbf{S}}_{j'} + J'' \sum_{\langle i,j''\rangle} \hat{\textbf{S}}_i \cdot \hat{\textbf{S}}_{j''} \nonumber \\* 
		& & + J_r \sum_{\langle i,j,k,l \rangle} \left[ (\hat{\textbf{S}}_i \cdot \hat{\textbf{S}}_j)(\hat{\textbf{S}}_k\cdot \hat{\textbf{S}}_l) + (\hat{\textbf{S}}_i \cdot \hat{\textbf{S}}_l)(\hat{\textbf{S}}_k \cdot \hat{\textbf{S}}_j)  \right. \nonumber \\ 
        & & \left. - (\hat{\textbf{S}}_i \cdot \hat{\textbf{S}}_k)(\hat{\textbf{S}}_j \cdot \hat{\textbf{S}}_l)\right] \label{eq:H4} .
	\end{eqnarray}
Here, $J$, $J'$ and $J''$ are exchange constants for first, second and third nearest neighbor couplings, respectively. $J_r$ (in some literature denoted $2K$\cite{majumdar12}) describes the ring exchange coupling that quantifies virtual circular currents. By performing the projection to single occupancies we make a truncation error that depends on the value of $t/U$. The coupling constants can be expressed in terms of the Hubbard constants $t$ and $U$ as:
    \begin{eqnarray}
	    J  = & \hspace{1pt} 4 t^2/U - 24t^4/U^3 \nonumber \\*
        J' = & J'' =  4t^4/U^3 \nonumber \\*
        J_r= & 80 t^4/U^3 \label{eq:couplingstrengths}
    \end{eqnarray}
Thus the ring exchange constant is always 20 times larger than both the second- and the third-neighbor Heisenberg exchange constant. This parametrization is based on a physical picture, where the electrons can only make jumps to nearest neighbor sites. The exchange process behind the second (and third) neighbor couplings therefore involves four jumps in total. All exchange processes involved in the perturbation are illustrated in Fig. \ref{fig:ringex}. It is worth noticing that the coupling strengths in Eq.~(\ref{eq:couplingstrengths}) all have the same sign, which means that the second- and third-nearest neighbors will be a source of frustration. Likewise, an increase in the ring exchange coupling will also result in increased frustration in the system. This is evident from the fact that a standard two-sublattice N\'{e}el order induces a negative energy contribution from the Heisenberg term and a positive contribution from the REM term.

The mean field energy per site of the REM can be calculated as:
   	\begin{eqnarray}
        E & = & 2 J (\textbf{S}_\textbf{A} \cdot \textbf{S}_\textbf{B}) + 2 J' (\textbf{S}_\textbf{A} \cdot \textbf{S}_\textbf{A}) + 2 J'' (\textbf{S}_\textbf{A} \cdot \textbf{S}_\textbf{A}) \\ \nonumber
        & & + J_r \left[ (\textbf{S}_\textbf{A} \cdot \textbf{S}_\textbf{B})  (\textbf{S}_\textbf{A} \cdot \textbf{S}_\textbf{B}) +  (\textbf{S}_\textbf{A} \cdot \textbf{S}_\textbf{B})  (\textbf{S}_\textbf{A} \cdot \textbf{S}_\textbf{B}) \right. \\ \nonumber
        & & \left. -  (\textbf{S}_\textbf{A} \cdot \textbf{S}_\textbf{A})  (\textbf{S}_\textbf{B} \cdot \textbf{S}_\textbf{B})\right]
    \end{eqnarray}
where \textbf{A} and \textbf{B} refer to two oppositely aligned ferromagnetic sublattices, which together defines a classical N\'eel state. Assuming that the spins are rotated an angle $\theta$ away from their perfect antiparallel alignment, and by employing the Hubbard-parametrization defined in Eq. (\ref{eq:couplingstrengths}), the following classical energy can be derived:
    \begin{equation}
        E = \frac{J_r}{8} \cos^2 \theta - \frac{1}{2} \cos \theta - \frac{J_r}{80}
    \end{equation}
    where we have set the value of $J$ to 1. Minimizing this expression with respect to $\theta$ yields:
    \begin{equation}
        \cos \theta = \frac{2}{J_r}
    \end{equation}
This expression has no solution for $J_r < 2$, meaning that N\'eel ordering is the classical ground state in this $J_r$-range. For $J_r > 2$, however, the ground ground state is characterized by two anti-aligned sublattices rotated by a finite angle $\theta$. This has implications for the linear spin wave results presented in section \ref{sec:LSW}, because these results have been derived under the assumption that the classical ground state is the  N\'eel state and are therefore not expected to give meaningful results for $J_r/J > 2$.

 \section{Linear spin-wave theory \label{sec:LSW}}
In linear spin-wave (LSW) theory the ground state is assumed to be the classical N\'eel ground state with opposite spins at neighboring sites of the square lattice. Appropriately, the spins of the ring exchange Hamiltonian are written in the Dyson-Maleev representation, with neighboring spins belonging to two  coordinate systems, $A$ and $B$, of oppositely aligned spins:
    \begin{align}
	    \hat{s}^+_{Ai} &	 = \sqrt{2 s} \left[ \hat{a}_i - \frac{\hat{a}_i^\dagger \hat{a}_i \hat{a}_i}{2s}\right], \hspace*{0.5cm} \hat{s}^-_{Ai} = \sqrt{2 s}\hat{a}_i^\dagger \label{eq:DysonMaleev1}\\
		\hat{s}^+_{Bj} &	 = \sqrt{2 s} \left[ \hat{b}_j^\dagger - \frac{\hat{b}_j^\dagger \hat{b}_j^\dagger \hat{b}_j}{2 s}\right], \hspace*{0.5cm} \hat{s}^-_{Bj} = \sqrt{2 s}\hat{b}_j \\
        \hat{s}^z_{Ai} & = s - \hat{a}_i^\dagger \hat{a}_i, \hspace*{0.5cm} \hat{s}^z_{Bj} = -s + \hat{b}_j^\dagger \hat{b}_j \label{eq:DysonMaleev}
    \end{align}
where $\hat{a}_i^\dagger$ and $\hat{b}_j^\dagger$ are the creation operators of up-spins on site $i$ and down-spins on site $j$, respectively. Following this transformation, $\hat{H}^{(4)}$ in Eq.~(\ref{eq:H4}) is diagonalized via a Bogoliubov transformation in analogy with Ref. \onlinecite{majumdar12}. This leads to the dispersion relation\cite{coldea2001, katanin2002}:
    \begin{align}
		\hbar \omega_\textbf{q} & = 2 s \left(l_\textbf{q}^2 + m_\textbf{q}^2 \right)\left( 2 J + J' \gamma_{1, \textbf{q}} + J'' \gamma_{2, \textbf{q}} \right. \nonumber\\
        & \hspace*{0.5cm} \left.  - J_r s^2\gamma_{3, \textbf{q}}  \right) + 4 s l_\textbf{q} m_\textbf{q} \gamma_{4, \textbf{q}} ( J  - 2 s^2 J_r) \label{eq:ringdisp}
	\end{align}
where $\gamma_{1, \textbf{q}}$, $\gamma_{2, \textbf{q}}$, $\gamma_{3, \textbf{q}}$ and $\gamma_{4, \textbf{q}}$ are trigonometric functions defined as:
    \begin{align}
    	\gamma_{1, \textbf{q}} & = \cos (q_x + q_y) + \cos( q_x - q_y) - 2, \\
        \gamma_{2, \textbf{q}} & = \cos(2 q_x) + \cos(2 q_y) - 2, \\
        \gamma_{3, \textbf{q}} & = \cos(q_x + q_y) + \cos ( q_x - q_y) + 2, \\
        \gamma_{4, \textbf{q}} & = \cos(q_x) + \cos(q_y).
    \end{align}
The Bogoliubov coefficients $l_\textbf{q}$ and $m_\textbf{q}$ are given by:
	\begin{align}
		l_\textbf{q}^2 & = \frac{1}{2} + \frac{\sqrt{x_\textbf{q}^2 - 4 x_\textbf{q} z_{\textbf{q}}}}{2(x_\textbf{q} - 4 z_\textbf{q})}, \nonumber \\
        m_\textbf{q} & = -\text{sign}(\gamma_{4, \textbf{q}})\sqrt{l_\textbf{q}^2 - 1}. \label{eq:ringbogo}
	\end{align}
with $x_\textbf{q}$ and $z_\textbf{q}$  defined as:
	\begin{align}
		x_\textbf{q} & = [8 J s + 4 J' s \gamma_{1, \textbf{q}} + 4 J'' s \gamma_{2, \textbf{q}}  - 4 J_r s^3 \gamma_{3, \textbf{q}}]^2 \\
		z_\textbf{q} & = (\gamma_{4, \textbf{q}})^2(4 J_r s^3 - 2 J s)^2
	\end{align}
If the third nearest neighbor coupling constant $J''$ is set to be 0, the dispersion relation from Ref. \onlinecite{majumdar12} is recovered. On the other hand, if one wishes to retain the Hubbard parametrization, the ratio 1:1:20 should be kept between $J'$, $J''$ and $J_r$. The overall strength of the Hubbard ring exchange coupling can therefore be characterized by a single value, $J_r/J$.
        
A LSW derivation of the the dynamic correlation function at momentum $\textbf{q}$ and energy $\omega$ ($\hbar=1$) starts from the definition\cite{dagotto94}:
	\begin{equation}
		S^{\alpha \beta} (\textbf{q},\omega) = \frac{1}{2 \pi} \int_{- \infty}^\infty dt  e^{- i \omega t} \sum_{l} e^{i \textbf{q} \cdot \textbf{r}_l} \langle \hat{s}_0^\alpha(0) \hat{s}_l^\beta (t)\rangle .
	\end{equation}
Here $\hat s_l^\alpha(t)$ denotes the spin component $\alpha \in \{x,y,z\}$ at site $l$ and time $t$. In the present study, we limit ourselves to zero external field, and thus the three diagonal parts of the dynamic correlation function will be equal, ${S^{xx}(\textbf{q},\omega) = S^{yy}(\textbf{q},\omega) = S^{zz}(\textbf{q},\omega)}$. We calculate only the longitudinal correlation function corresponding to ${\alpha =\beta = z}$. 

As in the derivation of Eq.~(\ref{eq:ringdisp}), we assume two oppositely aligned sublattices of spins and make use of the Dyson-Maleev representation from Eqs. (\ref{eq:DysonMaleev1}-\ref{eq:DysonMaleev}). Reusing the Bogoliubov coefficients from Eq. (\ref{eq:ringbogo}), the static structure factor is written as: 
  \begin{align}
    	S^{zz}(\textbf{q}, t = 0) & = \int_{-\infty}^\infty d \omega S^{zz}(\textbf{q}, \omega ) \nonumber \\
        & = \frac{s}{2} \left( l_\textbf{q}^2 + m_\textbf{q}^2 + 2 l_\textbf{q} m_\textbf{q}\right) \label{eq:strucSXX}
    \end{align}
which provides a theoretical base for comparisons with both numerical and experimental neutron scattering results. Due to quantum fluctuations, the mean eigenvalue of the spin-z operator will differ from the classical value of $\frac{1}{2}$. The deviation of the expectation value of the spin-z operator from s = $\frac{1}{2}$ is defined as:
    \begin{equation}
    	\delta \langle s^z \rangle = s - \langle s_i^z \rangle = \langle a_i^\dagger a_i \rangle
    \end{equation}
By Fourier transformation and inserting the magnon operators, the following expression is obtained:
   \begin{align}
   		\delta \langle s^z\rangle & = \frac{1}{N} \sum_\textbf{q} \langle a_\textbf{q}^\dagger a_\textbf{q} \rangle \nonumber \\
        & = \frac{1}{N} \left[l^2_\textbf{q} n_{\textbf{q},0} + m^2_\textbf{q} (n_{\textbf{q},1} + 1)\right] \nonumber \\
        & = \frac{1}{N} \sum_\textbf{q} m_\textbf{q}^2  , \hspace*{0.5cm} T = 0 \, {\rm K} \label{eq:szcor}
   \end{align}
   \begin{figure}
   	\centering
    \includegraphics[width = 7.74cm]{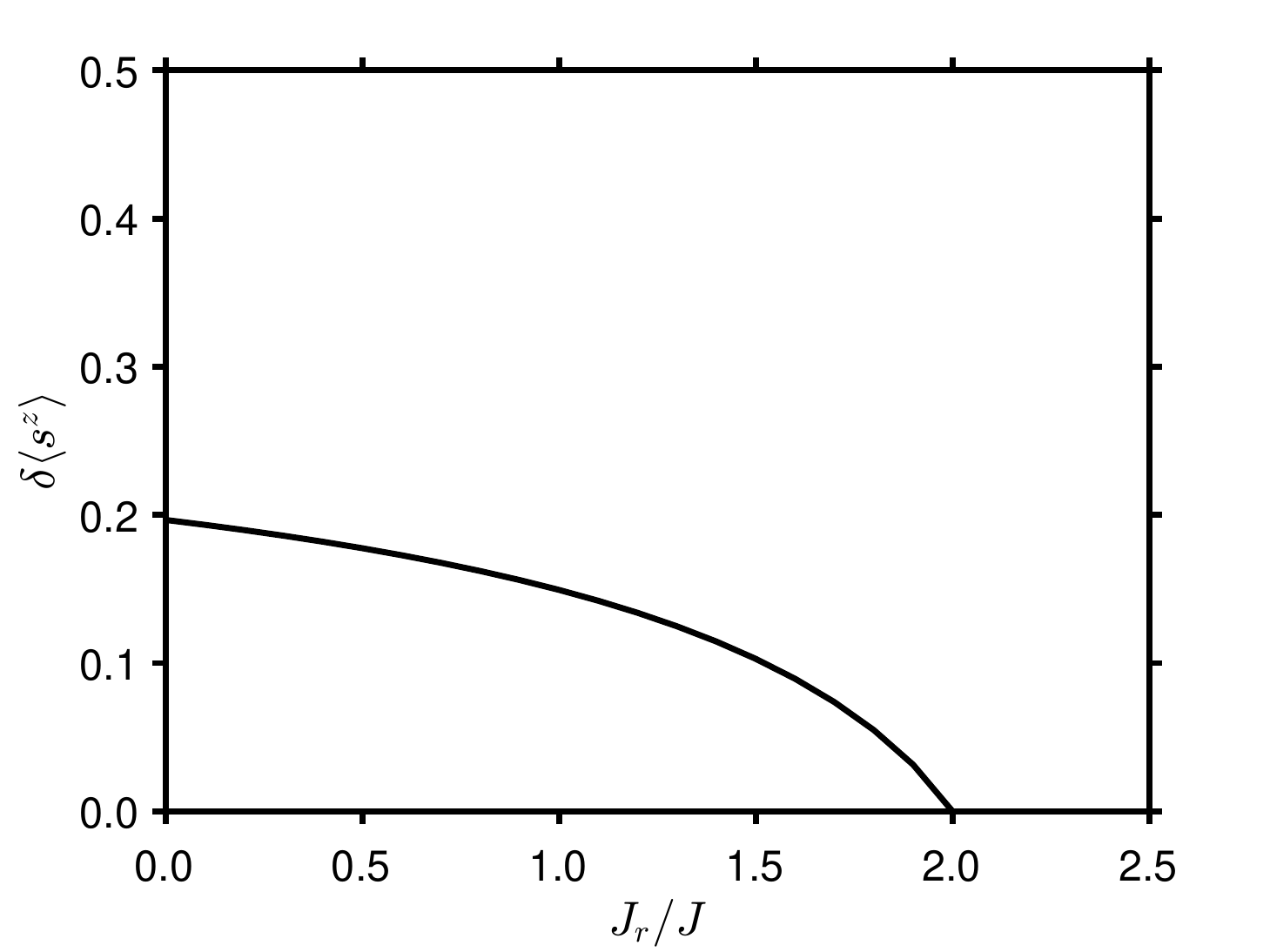}
    \caption{LSW deviation of the staggered magnetization from $1/2$ as a function of the ring coupling parameter. }
    \label{fig:quantumcorrection}
   \end{figure} 
where $n_{\textbf{q},0}$ and $n_{\textbf{q},1}$  refer to the Bose function for the two branching magnon modes. Thus the last equal sign is true at zero temperature only. 

Fig.~\ref{fig:quantumcorrection} shows the calculated LSW deviation of the mean staggered magnetization, $ \delta \langle s^z \rangle$, as a function of $J_r/J$. We see that the staggered magnetization counterintuitively reaches its maximum value of $1/2$ at the critical point $J_r/J = 2$. This is explained by the fact that first order quantum corrections exactly cancel at this point \cite{chubukov1992}. However, Ref.~\onlinecite{majumdar12} pointed out that higher order quantum corrections are indeed important around this value.

\section{Exact diagonalization method \label{sec:methods}}

In the present study, exact diagonalization of the spin-$1/2$ Hubbard-parametrized ring exchange model has been performed with the RLexact software package.\cite{lefmann} Spin clusters with periodic boundary conditions of size N = 16, 18, 20, 26, and 32 have been employed for the calculations. Despite claims of the opposite in Ref.~\onlinecite{chubukov1992}, we find no reason to limit the ED studies to square plaquettes of $n \times n$ spins with $n$ even.

\subsection{The choice of a symmetric basis}
The dimension of the matrices to be diagonalized can be greatly reduced by choosing a basis that will bring the REM Hamiltonian on a block-diagonal form. RLexact makes use of two symmetries of the REM, namely conservation of the total magnetization and conservation of lattice momentum.  By applying the magnetization symmetry operator, $\hat{s}^z$, the Hamiltonian is block-diagonalized into $N$+1 $m$-invariant subspaces, $m$ being the eigenvalue of $\hat{s}^z$. The momentum symmetry is present because of imposed periodic boundary conditions in two dimensions. The eigenvalues of the horizontal and vertical translation operators, $\hat{T}_x$ and $\hat{T}_y$, are defined from the eigenvalue problem:
	\begin{equation}
		\hat{T}_x \hat{T}_y | \Psi \rangle = e^{-i q_x} e^{-i q_y} | \Psi \rangle, \label{eq:eig}
		\end{equation}	 
		where $|\Psi\rangle$ is a spin state, and $q_x$ and $q_y$ are components of the momentum vector $\textbf{q}$. Given a spin system with $N_x$ spins along the $x$-direction and $N_y$ spins along the $y$-direction, application of the horizontal and vertical translation operators $N_x$ and $N_y$ times respectively must bring a state back to itself. As a result, the following relations hold:
		\begin{equation}
		q_x = \frac{2 \pi}{N_x} k_x, \hspace*{1cm} q_y = \frac{2 \pi}{N_y} k_y, \label{eq:momentum}
	\end{equation}
	where $k_x$  and $k_y$ are integers. These expressions underline the discrete nature of the numerical momentum-vector, which is caused by the finite size of the investigated clusters. Larger cluster sizes corresponds to a denser sampling of reciprocal space. The translation operator eigenstates, forming the basis for the exact diagonalization procedure, are constructed as superpositions of unique Ising states \cite{sandvik10}:
	\begin{equation}
		|u_n, m, q_x q_y \rangle = \frac{1}{\sqrt{N_n}} \sum_{n_x = 0}^{N_x-1} \sum_{n_y = 0}^{N_y-1} e^{i (q_x n_x + q_y n_y)} \hat{T}^{n_x}_x \hat{T}_{y}^{n_y} |u_n \rangle \label{eq:momentumstate}.
	\end{equation}	
Here, $n_x$	and $n_y$ are the number of times the translation operators are applied to the unique Ising state $|u_n\rangle$. A set of unique states is defined as a set of states that cannot be brought into one another via any combination of translation operations. 
    
    \subsection{The Lanczos diagonalization method \label{subsec:lanxmethod}}
   
\begin{figure}
		\centering
        \includegraphics[width = 7.74cm]{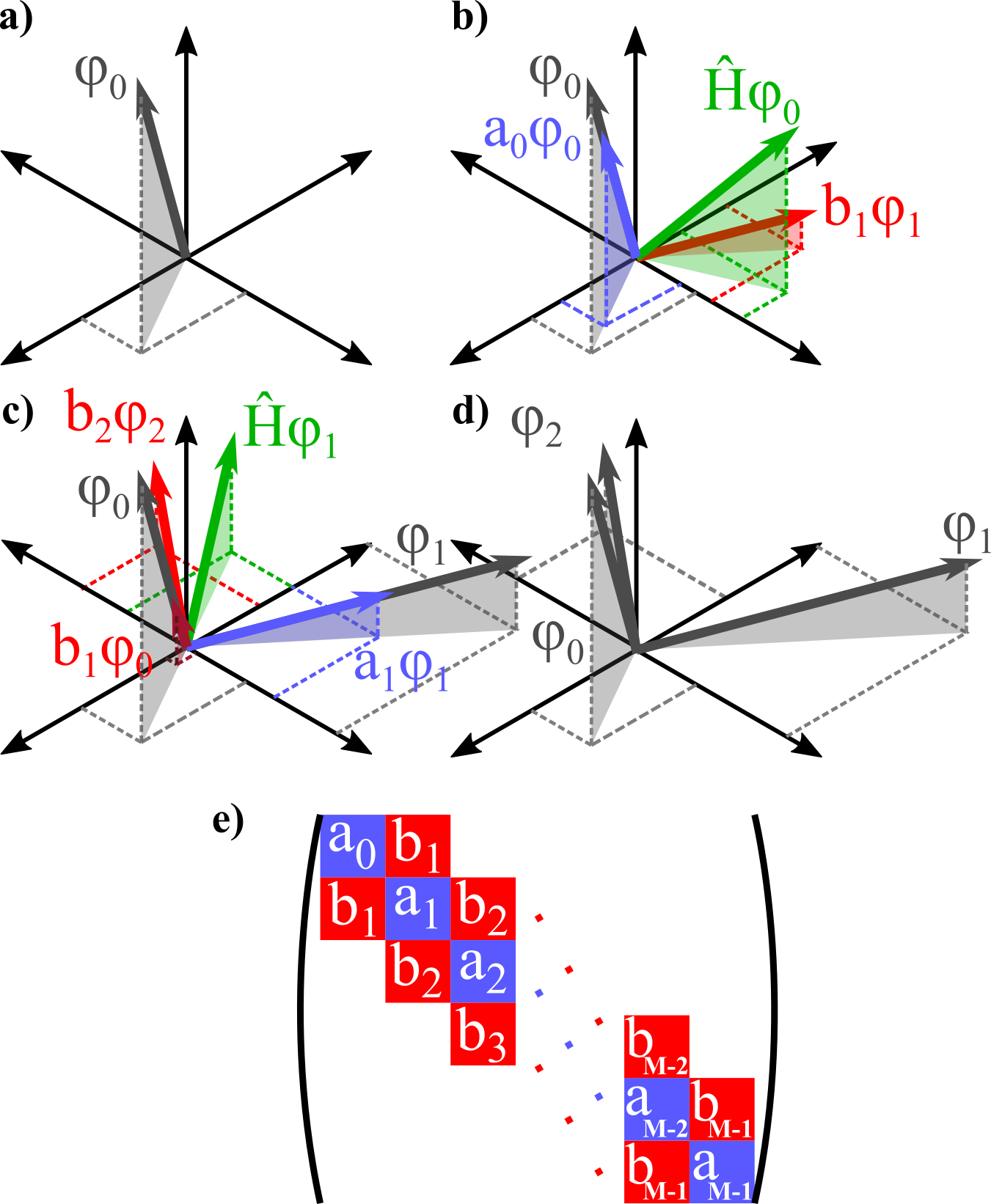}
          \caption{Initial steps of the Lanczos process\cite{Prelovsek13}. \textbf{a)} A random ''seed`` vector, $|\phi_0\rangle$, is used as the first state of the Krylov basis. \textbf{b)} The Hamiltonian is applied to $|\phi_0\rangle$ and components parallel and orthogonal to $|\phi_0\rangle$ are identified. \textbf{c)} The Hamiltonian is applied to $|\phi_1 \rangle$ and the same procedure is followed; a component of $\hat{H}|\phi_1 \rangle$ orthogonal to both $|\phi_1\rangle$ and $|\phi_0\rangle$ is identified as well as two components parallel to $|\phi_1\rangle$ and $|\phi_0 \rangle$, respectively. \textbf{d)} The mutually orthogonal states $|\phi_0\rangle$, $|\phi_1\rangle$, and $|\phi_2\rangle$ make up the first states of the Krylov space. \textbf{e)} A Hamiltonian projected onto this subspace will take a tridiagonal form.}
          \label{fig:lanczos}
    \end{figure}

Once block-diagonalized, the dimensions are further reduced via the Lanczos algorithm,\cite{lanczos50} which projects a given $(L \times L)$ Hamiltonian onto a smaller $(M \times M)$ Krylov subspace. The workings of the Lanczos algorithm is illustrated in Fig. \ref{fig:lanczos}. A tridiagonal matrix is constructed from a random starting seed, $|\phi_0\rangle$, by repeatedly applying the Hamiltonian to the Krylov eigenstates and determining components parallel and perpendicular to existing eigenstates:
    \begin{equation}
    	\hat{H} | \phi_i \rangle = b_i |\phi_{i-1}\rangle + a_i | \phi_{i} \rangle + b_{i+1} | \phi_{i+1} \rangle
    \end{equation}
here $|\phi_i\rangle$ and $|\phi_{i-1}\rangle$ are existing eigenstates of the Krylov subspace, and $| \phi_{i+1} \rangle$ is an eigenstate constructed such that it is orthogonal to them both. The parameters $a_i$, $b_i$ and $b_{i-1}$ are real constants that are chosen such that there is no overlap between eigenstates. This way of constructing eigenstates ensures that each newly generated eigenstate, $|\phi_{i+1}\rangle $, is orthogonal to every previous identified  eigenstate in the Krylov space, as has been proven by induction.\cite{sandvik10} An analysis of the accuracy and convergence properties of the Lanczos method has been performed in Ref. \onlinecite{paige80}. The extremal eigenvalues of the generated trigonal matrix are known to converge very quickly towards the actual extremal eigenvalues of the Hamiltonian, while the interior eigenvalues may be less accurate.
    
\begin{figure}[t]
    	\centering
        \includegraphics[width = 8.6cm]{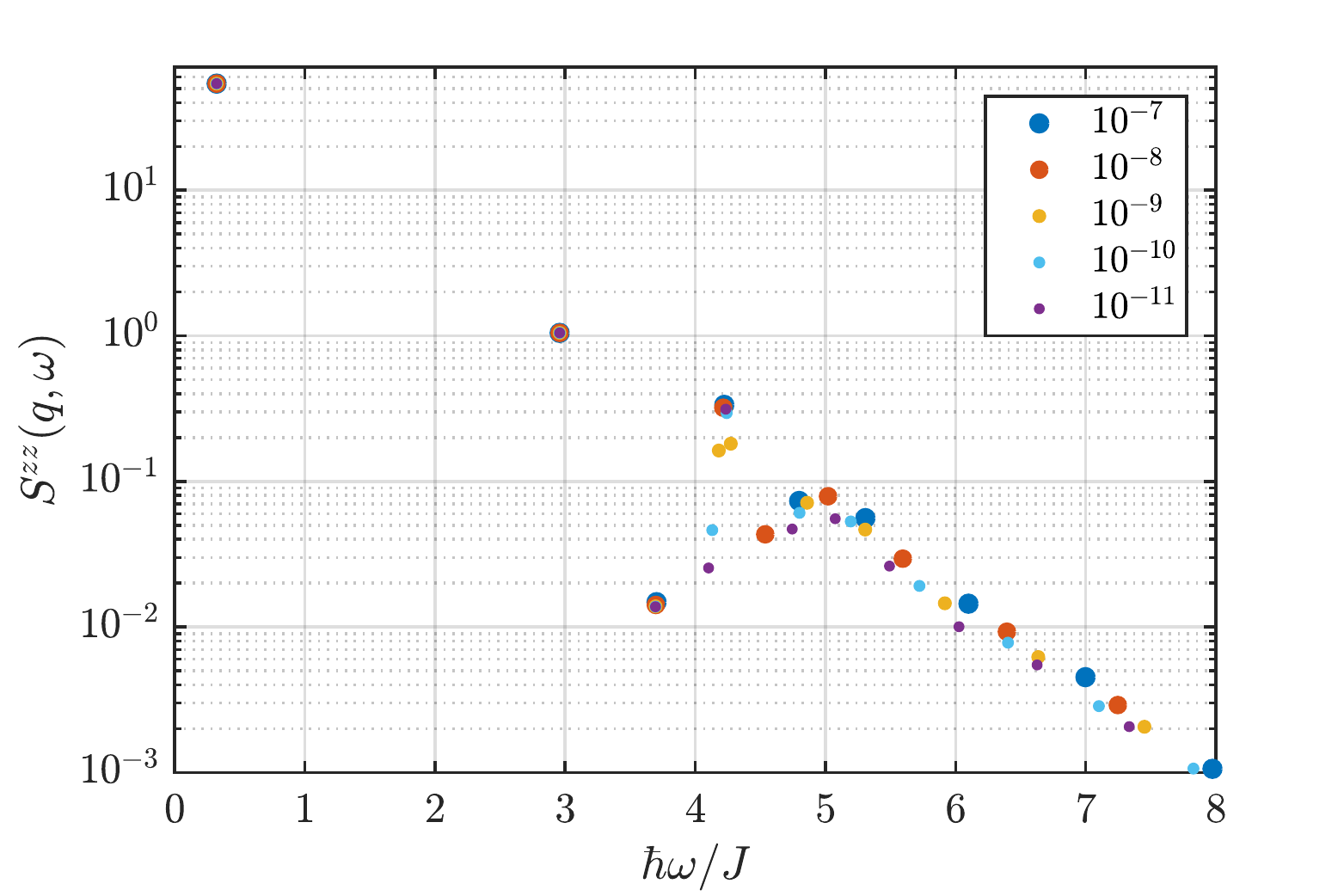}
        \caption{Dynamical correlation function, $S^{zz}(\boldsymbol{q},\omega)$ vs.~excitation energies at $\boldsymbol{q} = (\pi,\pi)$ of the four-spin ring exchange model on a square lattice with $N=32$ spins and $J_r/J=1$ as determined by ED with various Ritz values plotted on logarithmic scale.}
        \label{fig:ritzeffectRING}
    \end{figure}

RLexact uses the Ritz value, r$_Z$, as defined in Ref.~\onlinecite{ruhe00} as its convergence criteria. To further investigate the effect of r$_Z$ on especially the intermediate eigenvalues, we carried out a methodological study on the REM. 
This is shown in Fig.\ \ref{fig:ritzeffectRING}, where the excitation energies are plotted versus the obtained dynamical correlation function for the $\boldsymbol{q} = (\pi,\pi)$ subspace of the $N=32$ and $J_r/J=1$. {The data is plotted on a logarithmic axis, showing that the spectral weight is piled up in the low-energy part of the spectrum, as expected. The result is only weakly dependent on the choice of Ritz value in the range $10^{-7}-10^{-11}$, where the obtained excitation energies all fall onto the same trend line, with the difference that an increasing number of states appears for decreasing Ritz value, especially at higher energies $\hbar \omega/J > 4$. }
The discrepancy at larger energies arises because a lower convergence constant causes the Lanczos algorithm to run for more iterations and consequently to add more eigenstates to the Krylov basis. For the Ritz value of r$_Z$ = $10^{-7}$ a total of 21 excited states were found at $\boldsymbol{q} = (\pi,\pi)$, 29 states were found with r$_Z$ = $10^{-9}$, while 32 states were found with r$_Z$ = $10^{-11}$.\\
For the two lowest excitation energies, the values obtained with r$_Z=10^{-9}$ and r$_Z=10^{-11}$ differ  with less than machine precision, while the third excited energy differs with $10^{-4}$. Since a ratio of $J_r/J=1$ causes a relatively high frustration in the system, cases with lower $J_r/J$ will have better agreement between higher excited states for the same range of Ritz values. All presented ring exchange ED results were extracted with a Ritz value of $10^{-9}$. {We keep in mind that one should be cautious when interpreting REM data beyond the first couple of excitation energies. In addition, it is important to pay attention to possible finite size effects,\cite{bausch17} as we shall address in section \ref{sec:FSS}.}

\subsection{Dynamic correlation function \label{subsec:calccor}}
	
RLexact calculates the dynamic correlation factors using the Lehmann representation: 
	
    \begin{equation}
		S^{zz}(\textbf{q},\omega) = \sum_e M_e^{zz}(\textbf{q},\omega) \delta(\omega+E_0-E_e) \label{eq:Szz}
	\end{equation}
	
where $M_e^{zz}$ are the matrix elements calculated from the ground state $|0\rangle$ and a given excitation state $|\text{e} \rangle$:
	
    \begin{equation}
		M_e^{zz}(\textbf{q},\omega) = |\langle \text{e} | s_\textbf{q}^z | 0 \rangle|^2
        \label{eq:Mzz}
	\end{equation}
    
By using the state $s_\textbf{q}^z | 0 \rangle$ as the seed vector for the Lanczos algorithm, the inner products in Eq.~(\ref{eq:Mzz}) are found with a high degree of accuracy for the first few excited states because ED calculates the ground state of any given subspace to a very high precision. Additionally, the particular choice of seed favors states with a large value of the matrix element, and these are typically the lowest lying states.\\

Overall, the dynamic structure factors of any given system fulfills the following sum rule:\cite{lefmann}
\begin{equation}
\int_{-\infty}^\infty d\omega \sum_{\alpha,\textbf{q}} S^{\alpha \alpha}(\textbf{q}, \omega) = \sum_{\alpha,\textbf{q}} M^{\alpha \alpha} (\textbf{q}, \omega) = N S (S + 1) \label{eq:sumrule}
\end{equation}
meaning that even though different spin models may result in different spectral distributions, the overall sum of Eq. (\ref{eq:Mzz}) over all excitations will always be the same for a given system size.

\section{Numerical results\label{sec:disp}}
This section presents the RLexact calculated results based on the Hubbard-parametrized REM. Dynamical spin wave dispersions are first presented due to their applicability in the analysis of inelastic neutron scattering data. Thereafter, we show the static results. To investigate how well-suited numerical small cluster results are for the interpretation of experimental data involving orders of magnitudes more spins, a detailed discussion of finite size effects will follow.

We will focus on the high-symmetry wave vectors $\textbf{q}_\textbf{M} = (\pi, 0)$, $\textbf{q}_\textbf{S} = \left(\frac{\pi}{2}, \frac{\pi}{2} \right)$, and  $\textbf{q}_\textbf{X} = \left(\pi, \pi \right)$, because these wave vectors are well represented by the various system sizes  and because of the interesting physics reported at wave vectors between $\textbf{q}_\textbf{M}$ and $\textbf{q}_\textbf{S}$ in the unperturbed case \cite{piazza15,ourotherpaper}.

\subsection{Excitation spectra}
    
\begin{figure*}
	\centering
	\includegraphics[width=0.9\linewidth]{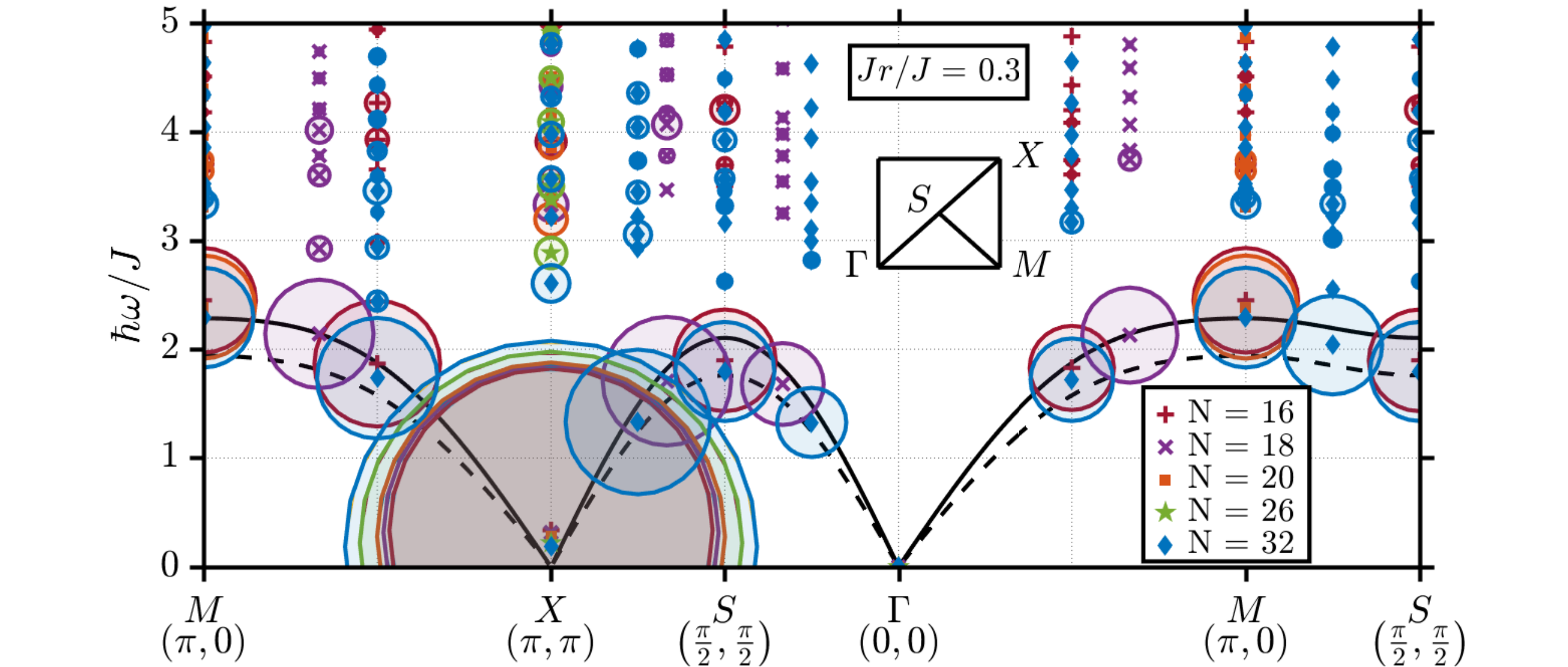}
    \includegraphics[width=0.9\linewidth]{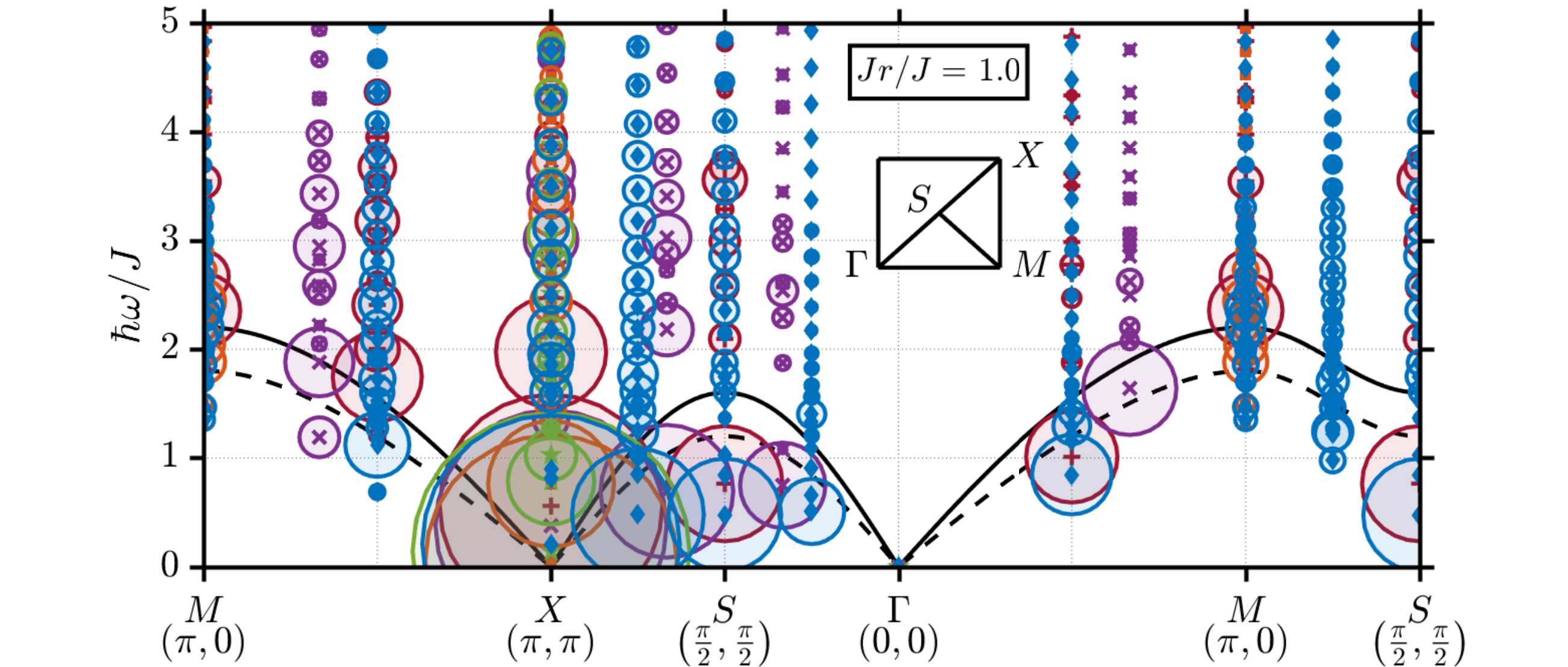}
	\caption{Numerical dispersion results for the ring exchange model with J$_r$/J = 0.3 and J$_r$/J = 1 as calculated with ED with 16-spin (red), 18-spin (purple), 20-spin (orange), 26-spin (green) and 32-spin (blue) spin clusters. The area of the circles around each excitation point is proportional to the dynamic correlation factor, Eq. (\ref{eq:Szz}). The solid and dashed lines are the spin wave dispersions as calculated through linear spin wave theory, Eq. (\ref{eq:ringdisp}), with and without a quantum renormalization factor, respectively.}
    \label{fig:disp}
\end{figure*}

ED excitation spectra with $J_r/J$ = 0.3 and 1.0 are presented in Fig. \ref{fig:disp} along with the LSW dispersion defined in Eq.~(\ref{eq:ringdisp}). In the $J_r/J= 0.3$ spectrum, the first excitations carry the most spectral weight and roughly follow the LSW-calculated dispersions. The dashed-line LSW dispersion has been calculated according to Eq.~(\ref{eq:ringdisp}), while the solid-line LSW dispersion additionally has been renormalized with a quantum correction factor, $Z_c(\textbf{q})$.\cite{bastienthesis, piazza}

Qualitatively the $J_r / J$ dispersion resembles the unperturbed ($J_r=0$) antiferromagnetic Heisenberg dispersion, with one exception at $\textbf{q}_\textbf{M}$.  In the unperturbed case, a characteristic dip was observed in the first excitation energy at $\textbf{q}_\textbf{M}$ when compared to $\textbf{q}_\textbf{S}$. This dip could not be replicated by LSW theory or numerically by systems with 16 or less spins, indicating a size-dependent quantum feature. \cite{ourotherpaper} The $J_r/J = 0.3$ dispersion, on the contrary, exhibits a larger first excitation energy at $\textbf{q}_\textbf{M}$ than at $\textbf{q}_\textbf{S}$. This is true both for the $N = 32$ \textit{and} the $N = 16$ systems. Additionally, the LSW dispersion is also able to replicate the behaviour. In the ED data, the $N = 16$ system exhibits a first excitation difference of $e_1(\textbf{q}_\textbf{M}) - e_1(\textbf{q}_\textbf{S}) = 0.36 J$, while  the $N = 32$ system shows a very similar difference of $0.28 J$. The appearance of enhancement of the  first excitation energy at $\textbf{q}_\textbf{M}$ is therefore not a size-driven quantum feature, but behaves qualitatively different compared to the dip in the unperturbed case. 
    
The dispersion spectrum of La$_2$CuO$_4$ could in Ref.~\onlinecite{majumdar12} be reproduced by LSW calculations with $1/S^2$ corrections and a (non-parametrized) ring exchange value of around $J_r / J \approx 0.3$. The first excitation energy at $\textbf{q}_\textbf{S}$ is $\sim 13 \%$ lower than at $\textbf{q}_\textbf{M}$ in the experimental La$_2$CuO$_4$ data\cite{headings}. ED data exhibit a dip of $22 \%$ for a $N = 32$ system and $23 \%$ for a $N = 16$ system, both with a ring exchange coupling of $Jr/J = 0.3$. As of now, the ED results therefore overestimates the dip, which either indicates finite-size effects or an overestimation of the ring exchange coupling. 

In this work, we also report the case of $J_r / J = 1 $, {\em i.e.}\ a much stronger ring exchange coupling than what has been observed experimentally, to test the limits of the model and pick out characteristic features of the REM. In the lower plot of Fig. \ref{fig:disp}, it is evident that there is now much less agreement between the first excitation ED results and the LSW calculations.

In the $J_r / J = 0.3$ spectrum, all first excitations consistently have higher energy than the LSW predictions from Eq. (\ref{eq:ringdisp}). At the same time, the larger system sizes also result in increasingly lower first excitation energies, which indicates a convergence towards the LSW dispersion as the number of spins is increased. Application of the quantum correction factor, $Z_c(\textbf{q})$, results in an overall increase of the predicted dispersion energies, causing the LSW dispersion to be overestimated compared to the ED results at certain wave vectors. A similar effect is seen in the $J_r/J = 1$ dispersion, where both LSW predictions overshoot at most wave vectors, in particular at $\textbf{q}_\textbf{S}$. This could indicate a softening of a mode at that wave vector, as a precursor for a (quantum) phase transition as $J_r/J$ increases.

Gutzwiller projections have in Ref.~\onlinecite{piazza15} been used to investigate the dispersion spectrum of a N\'{e}el ground state versus a resonating valence bond (RVB) ground state on the square lattice. The latter ground state is characterized by a continuum of spectral weight at $\textbf{q}_\textbf{M}$ in the spin wave dispersion. A similar continuum appears to emerge in the ring exchange dispersion when the coupling strength is strong enough, as is seen in the $J_r / J = 1$ dispersion where the density of states at each wave vector has increased significantly. Furthermore, the lowest lying excitations do no longer consistently contain the largest amount of spectral weight. Instead, the dispersion shows a general trend of shifting the spectral weight up in energy, which is especially apparent at $\textbf{q}_\textbf{M}$, but also seen to some extend at  $\textbf{q}_\textbf{X}$.
	
\begin{figure}[!h]
	\centering
    \includegraphics[width = 7.74cm]{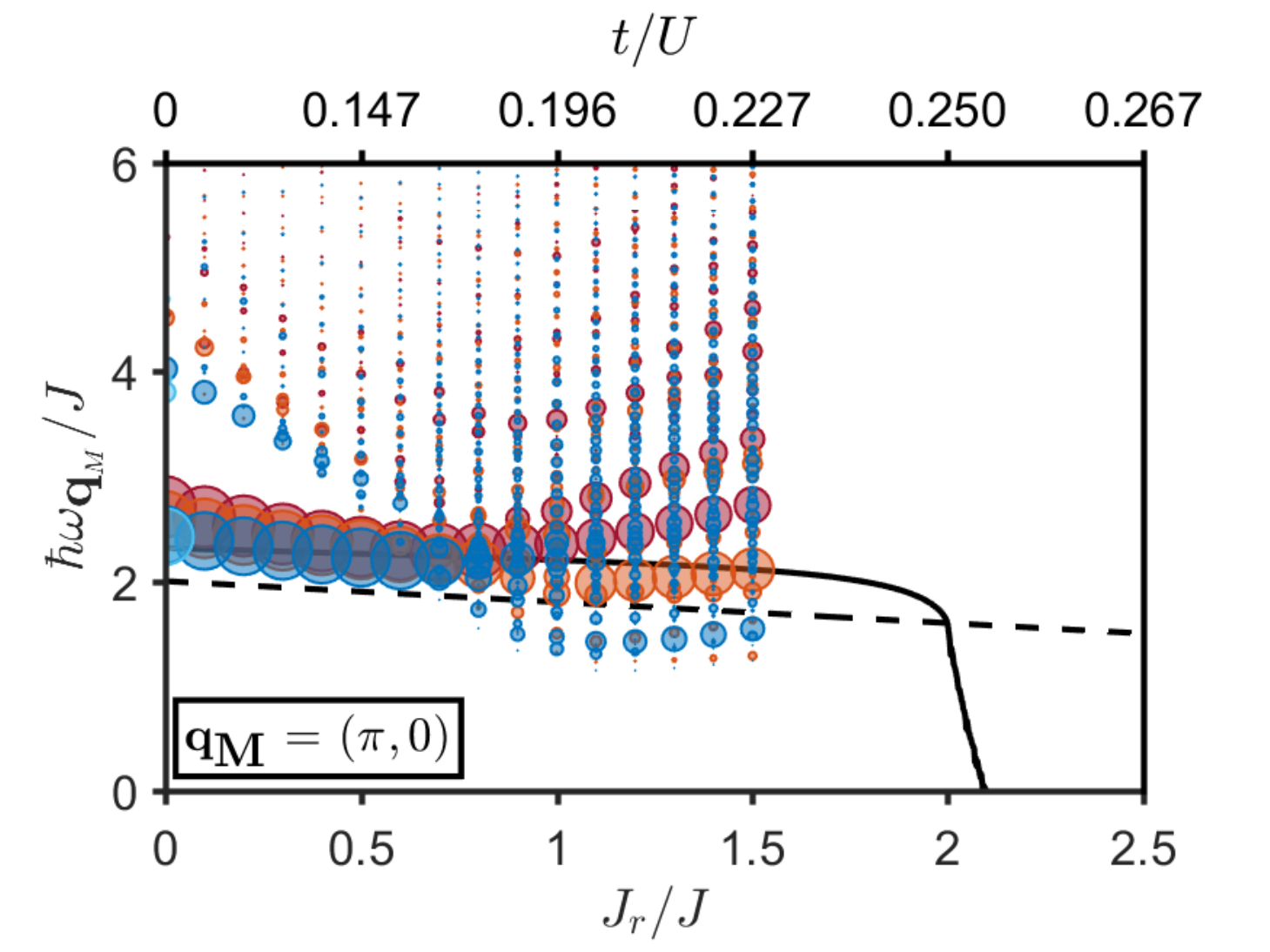}
    \includegraphics[width = 7.74cm]{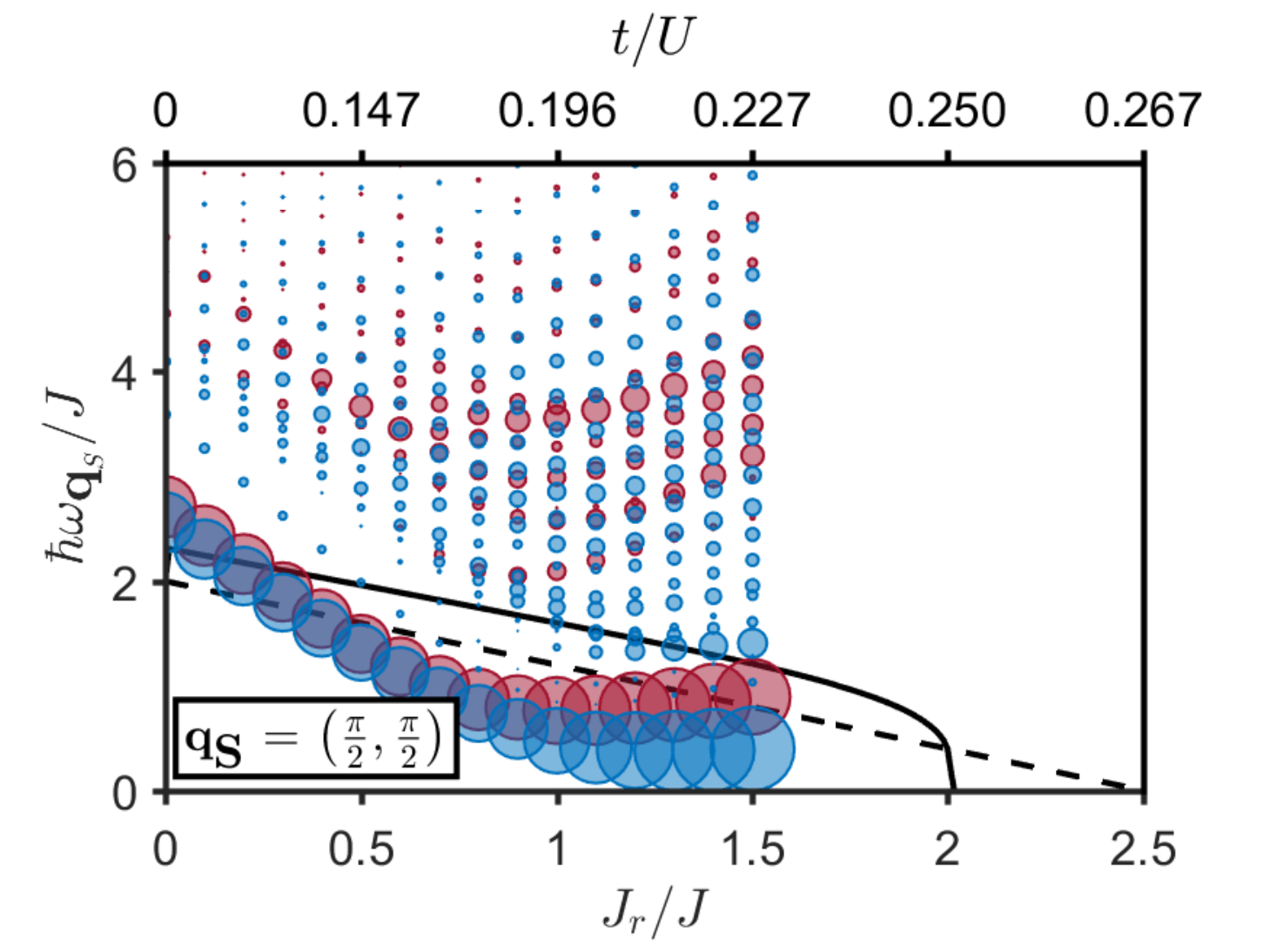}
    \includegraphics[width = 7.74cm]{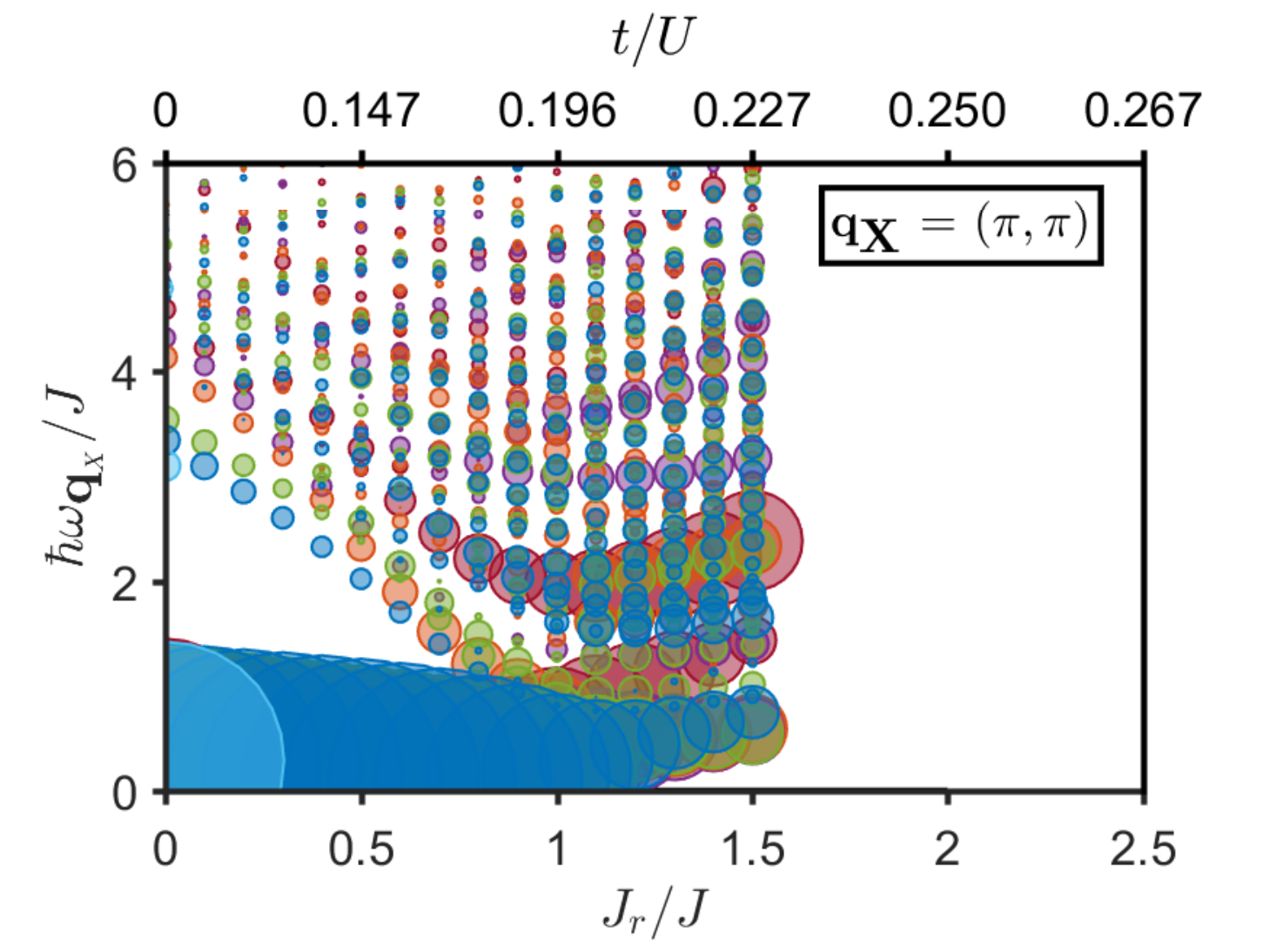}
    \caption{ED results for the excitation energies and $S^{zz}$ values at $\textbf{q}$ = $(\pi,0)$, $\left( \frac{\pi}{2}, \frac{\pi}{2} \right)$, $(\pi,\pi)$ as a function of the ring exchange coupling constant. Calculations have been carried out with systems of size N = 16 (red), 18 (purple), 20 (orange), 26 (green), 32 (dark blue), and 36 (light blue). The area of the circles are proportional to $S^{zz}(\textbf{q}, \hbar \omega)$. The dashed line is the LSW dispersion from Eq. (\ref{eq:ringdisp}) with no quantum correction, and the solid line is the same dispersion with a quantum correction factor applied. The value of the hopping parameter $t$ on the top axis has been calculated by assuming a value $U$ of 3.5 eV, which is the magnitude of the Coulomb potential in the Mott-insulator Sr$_2$CuO$_2$Cl$_2$\cite{hasan2000}. }
    \label{fig:dispszz}
\end{figure}

\begin{figure}
    \centering
    \includegraphics[width = 7.74cm]{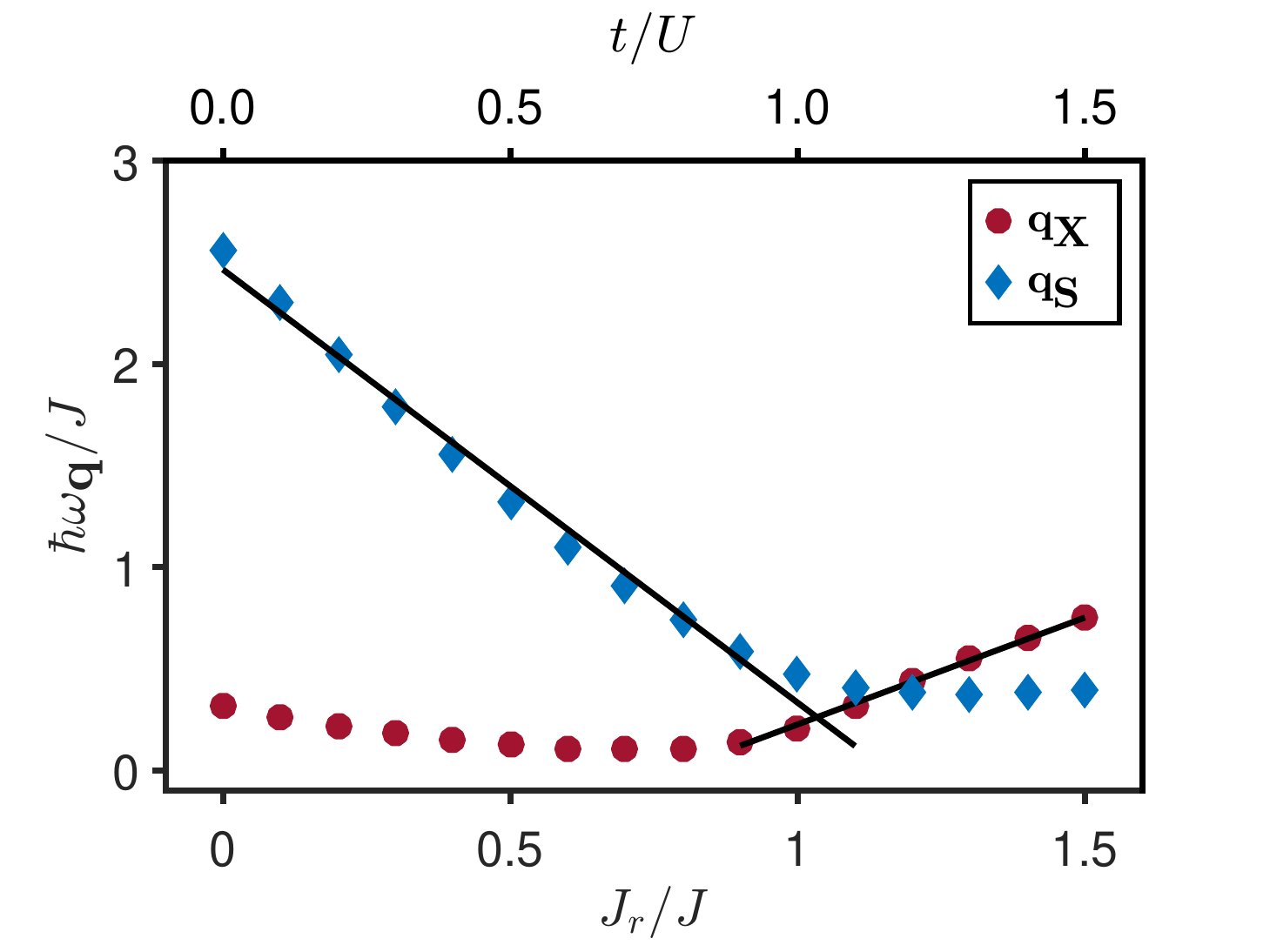}
    \caption{The energies of the excitation states at $\textbf{q}_\textbf{X}$ and $\textbf{q}_\textbf{S}$ with the largest dynamical structure factors. The data points portray the N = 32 ED data, as portrayed in Fig. \ref{fig:dispszz}. The solid lines are used as guides for the eyes and to highlight linear behaviour.}
    \label{disp:dispxands}
\end{figure}

A further investigation of the evolution of the excitations at specific wave vectors as a function of $J_r / J$ is carried out in Fig. \ref{fig:dispszz}. At $\textbf{q}_\textbf{M}$ we observe that that the gap between the first and second excitations decreases with increasing $J_r / J$. The number of excited states also increases, and the spectral weight is gradually shifted upwards in the excitation spectrum, as was seen in the case of $J_r / J = 1$ in Fig.~\ref{fig:disp}. For small system sizes, $N = 16$ and $N = 20$, the shift in spectral weight mostly happens from the first to the second excitation. On the other hand, at $J_r / J \geq 0.7$ the first excitation of the $N = 32$ spectrum seems to split up into a multitude of smaller poles with a more even distribution of spectral weight. At $\textbf{q}_\textbf{M}$, it is evident that the inclusion of the quantum correction, $Z_c(\textbf{q}$), drastically affects the ring exchange value, at which the LSW gap closes, in this case occurring at $J_r / J \approx 2.1$. This could imply changes in the magnetic order of the ground state, corresponding to the mean field phase transition at $J_r/J = 2$. 

At  $\textbf{q}_\textbf{S}$ a similar closing of the gap between the first and second excitation is observed, though the first poles still retain the most spectral weight even at strong $J_r / J$ couplings. We observe a general softening of the magnon mode with increased $J_r/J$ in both the ED and LSW calculations at $\textbf{q}_\textbf{S}$. The gap to the first excitation  is seen to close near to the mean field phase transition value, at $J_r/J \approx 2.0$, for the LSW calculations with the quantum renormalization factor, $Z_c(\textbf{q})$. The closing happens at a slightly higher value, $J_r/J \approx 2.5$,  without the quantum renormalization. In the ED calculations, the softening takes place at lower values of $J_r/J$, which we discuss below.

In the $\textbf{q}_\textbf{X} = (\pi, \pi)$ plot, we observe a low-lying mode for low values of $J_r/J$ as well as a closing of the gap between the first and second excitation taking place at $J_r/J \sim 1$. We also observe a significant redistribution of the spectral weight. Interestingly, it is also apparent that the lowest lying poles of the $N = 20$ and $N = 32$ systems switch places at $J_r / J \geq 1.3$, indicating the the excitation energy no longer extrapolates to 0 as one would expect from a symmetry-breaking N\'{e}el ordering (with ordering vector $\textbf{q}_\textbf{X}$) with an associated Goldstone mode. This is a further indication of the instability of the N\'{e}el phase, when strong ring exchange couplings are invoked.
    
Fig.\ \ref{disp:dispxands} shows the development of the gaps at $\textbf{q}_\textbf{S}$ and at $\textbf{q}_\textbf{X}$. We see that the gap closes almost linearly with $J_r/J$ for the former wave vector and opens almost linearly at the latter wave vector. These two behaviours happen almost at the same value of the ring exchange, $J_r/J = 1.0(1)$, indicating the possibility of a quantum phase transition between the N\'eel state and a state of ordering vector $\textbf{q}_\textbf{S}$ around this value of the ring exchange parameter. The detailed nature of the ground state $J_r/J > 1$ remains a topic for further investigation
    
Destabilization of the mean field N\'{e}el phase has already been documented in Heisenberg models with added ring exchange terms in geometries such as square lattices\cite{majumdar12}, triangular lattices\cite{motrunich05, holt14}, and four-leg triangular spin ladders\cite{block11}. The N\'{e}el phase usually gives way for a quantum spin liquid, which among other indicators is detected by its  excitation spectrum. Fractional spinon excitations result in an excitation continuum, as was observed in the aforementioned Gutzwiller-projection study.\cite{piazza15} A qualitative difference between the Gutzwiller dispersion spectrum and the ring exchange ED calculated dispersions in this study is that the Gutzwiller dispersion only contains a continuum at $\textbf{q}_\textbf{M}$. This difference may be caused by the used mean-field decoupling method, which for the RVB part only sums over nearest neighbors and does not contain an explicit ring exhange coupling.
    
\begin{figure*}
    	 \centering
         \includegraphics[width = 0.7 \linewidth]{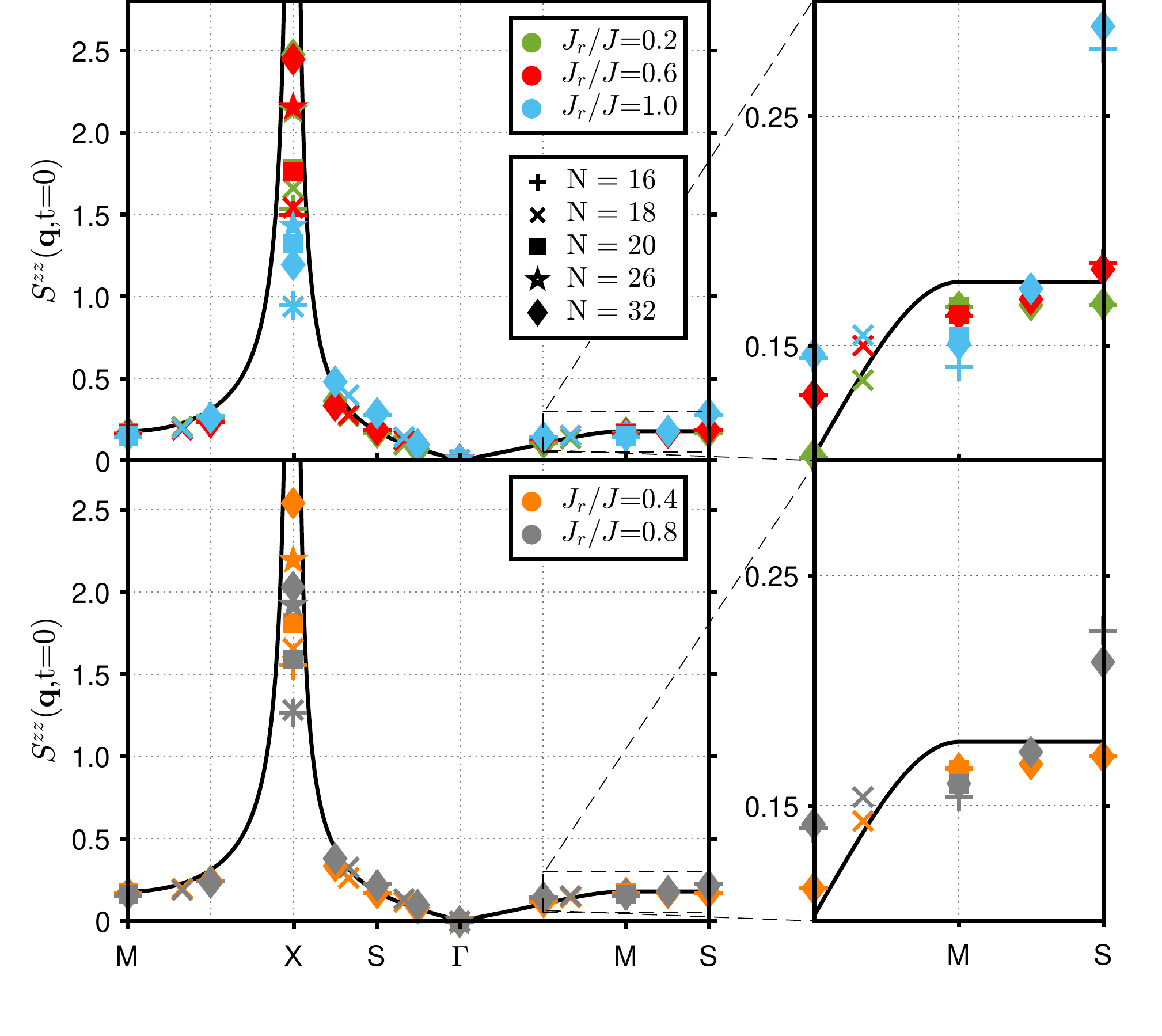}
         \caption{\textbf{Left}: Exact diagonalization results of the structure factor of the Hubbard-parametrized ring exchange model along the same cut through the Brillouin zone as used in Fig. \ref{fig:disp}. Different colours are used to separate data obtained with different coupling strengths, while the data symbol separate different system sizes. The solid black lines are the LSW structure factor, Eq. (\ref{eq:strucSXX}), normalized with respect to the sum rule in Eq. (\ref{eq:sumrule}). \textbf{Right}: A zoom-in of the behaviour of the structure factor along the direct path between $\textbf{q}_\textbf{M}$ and $\textbf{q}_\textbf{S}$. }
         \label{fig:struc}
    \end{figure*}

\subsection{Static structure factors}

The ED static structure factor is found by summing up the dynamic structure factors found via Eq.\ (\ref{eq:Mzz}). The results are shown in Fig. \ref{fig:struc}, where we show only the effect within the $(\pi,\pi)$ ordered phase, $J_r/J \leq 1$. As was the case for the dispersion result, the LSW results and ED structure factor results are less agreeable at strong ring exchange couplings, owing to stronger quantum fluctuations and a departure from the LSW assumed N\'{e}el order. The LSW dispersions are qualitatively the same between different $J_r / J$ coupling strengths, with the main difference being a renormalization factor related to the quantum correction of the sublattice magnetization, as derived in Eq. (\ref{eq:szcor}). The sum rule given in Eq. (\ref{eq:sumrule}) can be applied to the LSW structure factors by integrating numerically over the entire Brillouin zone:
    \begin{align}	
    	& \frac{1}{(2\pi)^2} \int_{-\infty}^\infty \int_\text{BZ} d \textbf{q} S^{zz}(\textbf{q}, \omega) =  \nonumber \\ 
        & \hspace*{0.75cm}\frac{1}{(2\pi)^2} \int_\text{BZ} d \textbf{q} S^{zz} (\textbf{q}, t = 0) = \frac{1}{4}
    \end{align}
    In the Heisenberg limit, $J_r / J = 0$, the left-hand side of the above expression is found to give 0.2113, which is 15~\% lower than the expected value of $\frac{1}{4}$. Upon changing the ring exchange coupling in the LSW structure factor to $J_r / J = 1.0$, an even lower value of 0.2013 is calculated. Thus, there is significant spectral weight associated with higher-order terms that have been excluded from the linear spin wave calculations. These higher order terms are found to be more integral to the REM with $U \gg t$ due to the more pronounced deviance from the sum rule at large $J_r/J$. If one normalizes the LSW structure factor with respect to the sum rule, a $J_r / J$-independent structure factor is found, as shown with solid lines in Fig.~\ref{fig:struc}.

      A clear $\textbf{q}$-dependent behaviour is seen in the ED structure factor results, when the $J_r / J$ coupling strength is varied. This is particularly evident at the wave vectors $\textbf{q}_\textbf{M}$ and $\textbf{q}_\textbf{S}$, as highlighted by the right plots of Fig. \ref{fig:struc}. The ring exchange coupling causes spectral weight to be shifted from $\textbf{q}_\textbf{M}$ to $\textbf{q}_\textbf{S}$. This is not an effect reflected in the LSW structure factors. Experimentally this behaviour has been observed in a neutron scattering study of La$_2$CuO$_4$, which resulted in a ring exchange description of the compound with parameters $J = 143(2)$~meV, $J' = J'' = 0.020(1) J$, and $J_r = 0.41(3) J$\ ~\cite{headings}. The study reports that the structure factor at $\textbf{q}_\textbf{M}$ was measured to be 50\% lower than at $\textbf{q}_\textbf{S}$. The ED results for the $J_r / J = 0.4$ system are unable the replicate this by reporting a difference of only $\approx 3 \%$. However, this discrepancy may be caused by finite-size effects, as will be discussed below, because a quantum Monte Carlo study managed to adequately describe the behaviour at $\textbf{q}_\textbf{M}$\cite{headings}. The experimental results are not completely out of line with the general trend of the ED results, since the difference between $S^{zz}(\textbf{q}_\textbf{M})$ and $S^{zz}(\textbf{q}_\textbf{S})$ increases as a function of $J_r/J$. With $J_r/J = 0.5$, a $7$\% difference is observed, while it is $12\%$ with $J_r/J = 0.6$. Thus a better agreement between experimental and ED static results is achieved if the $J_r / J$ value is adjusted to higher values. However, given that the agreement with the dynamical results are improved by lowering the $J_r / J$ value, it is likely that better agreement can be found with larger system sizes.

\subsection{Finite-size effects and extrapolations to the thermodynamic limit}
\label{sec:FSS}

\begin{figure}
	\centering
	\includegraphics[width = 8.6cm]{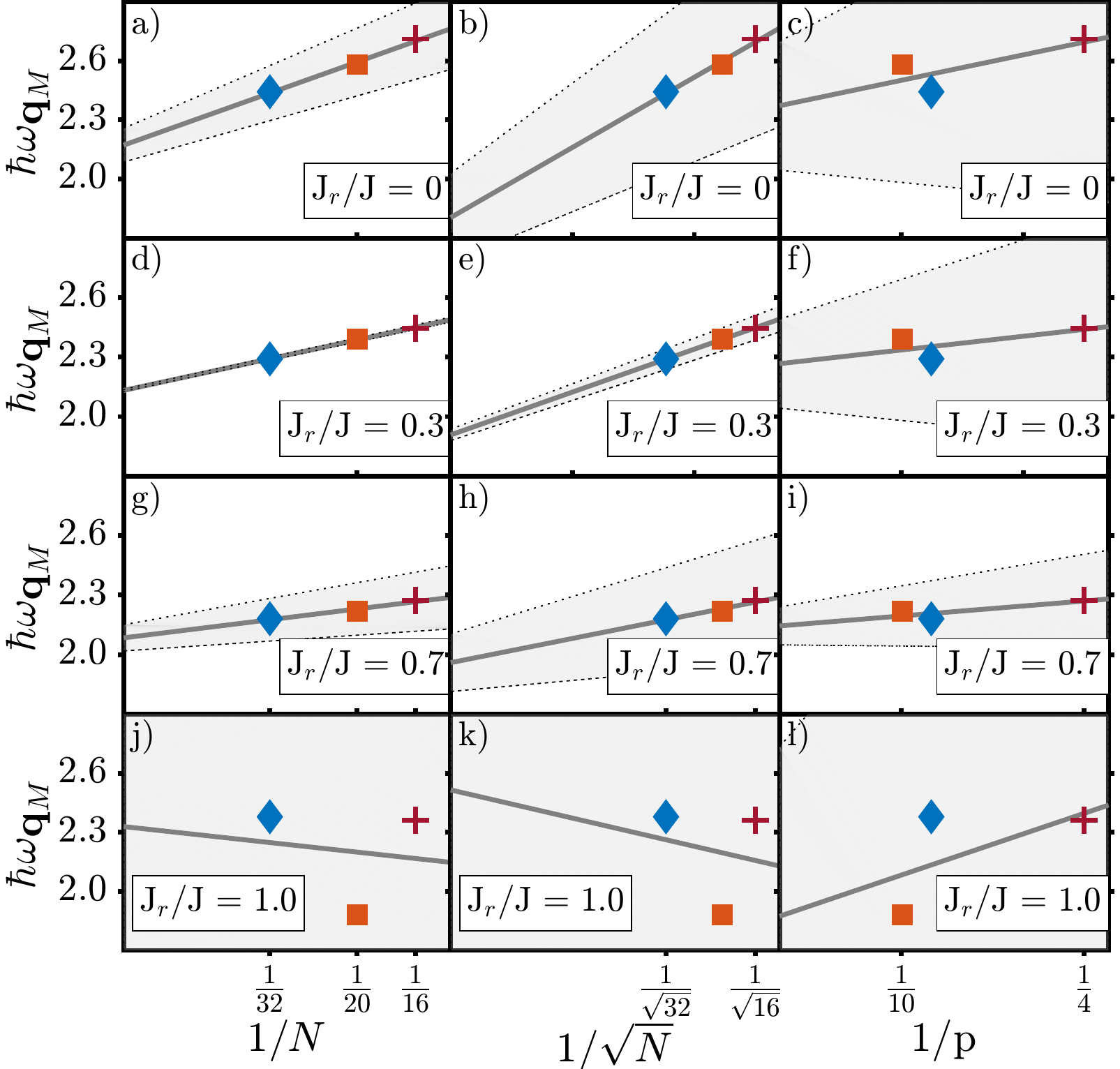}
    \caption{Different extrapolation methods for predicting the thermodynamic excitation energy of the first excited state with the largest value of $S^{zz}(\textbf{q}_M, \omega)$. The first column shows extrapolations based on the total system size, ${1/N \to 0}$, for selected $J_r / J$ values, the second column shows extrapolation based on the unit cell lengths ${1/\sqrt{N} = {1}/{L} \to 0}$, and the third column based on the periodicity of the systems, $\frac{1}{p} \to 0$. The solid grey lines are linear fits, while the shaded areas corresponds to fits with parameters within $\pm 1$ standard deviation of the grey line.} 
    \label{fig:extrappi0}
\end{figure}

\begin{figure}
	\centering
	\includegraphics[width = 8.6cm]{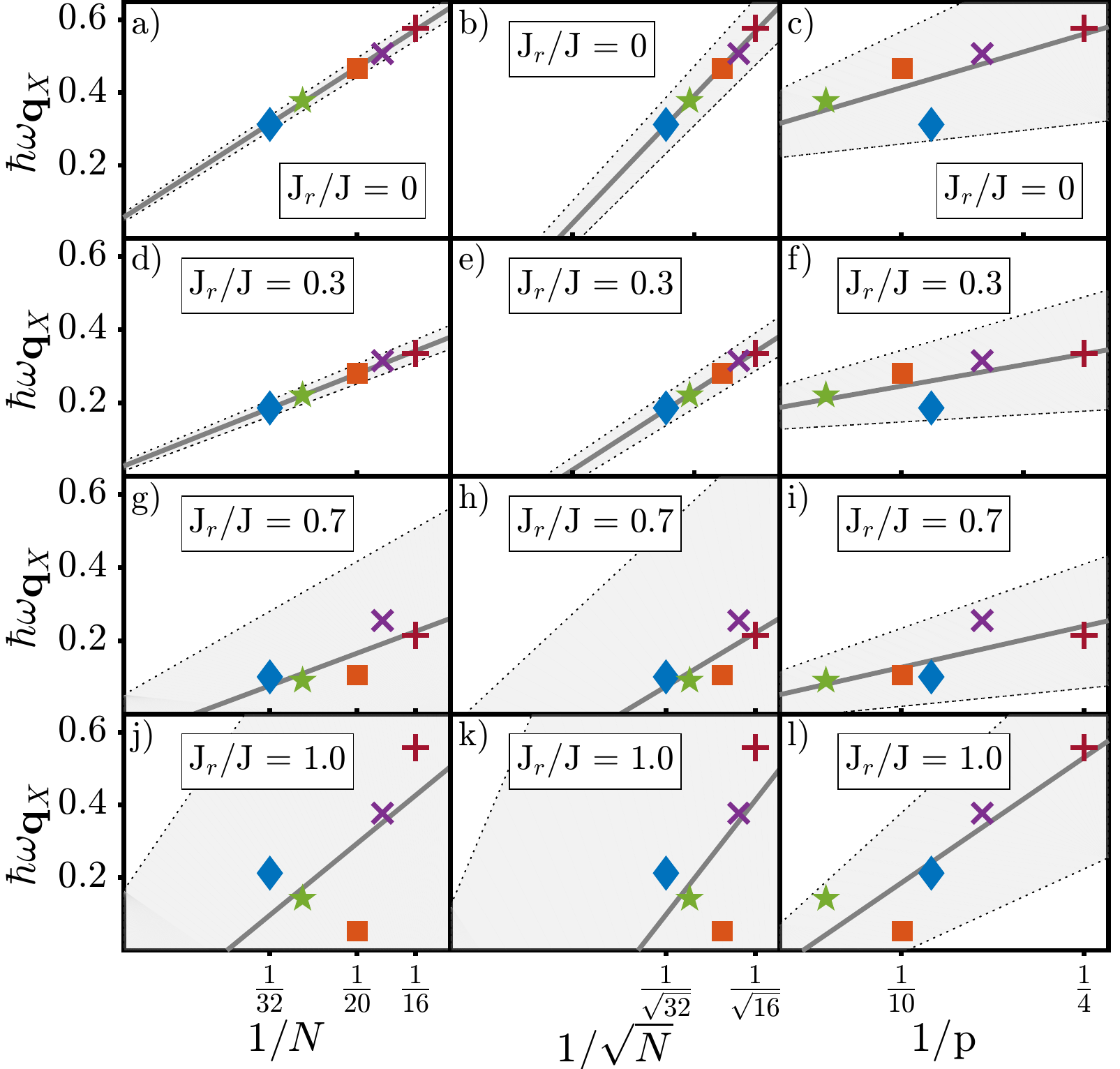}
    \caption{Different extrapolation methods for predicting the thermodynamic excitation energy of the first excited state with the largest value of $S^{zz}(\textbf{q}_X, \omega)$, similar to what has been done in Fig. \ref{fig:extrappi0}.}
    \label{fig:extrappipi}
\end{figure}

ED is biased towards small system sizes because of the rapidly increasing computational cost of the calculations with increasing number of spins. The ring exchange term further aggravates this problem, because it more than doubles the number of entries in the sparse matrix, thereby increasing computational time and limiting the largest system size of our calculations to $N = 32$. A prevailing way of interpreting numerical results based on small systems is to perform either a linear or square root extrapolation to the thermodynamic limit ($\frac{1}{N} \to 0$, $\frac{1}{\sqrt{N}} \to 0$). However, care should be taken in this approach due to the size-driven effect of certain quantum phase transitions, which may only appear at system sizes larger than those addressable by ED\cite{bausch17}.
        
In the past, extrapolations to the thermodynamic limit performed for the Heisenberg antiferromagnet on a square lattice have lead to extrapolated ED results that agree with quantum Monte Carlo results\cite{sandvik}. However, as was pointed out by L{\"u}scher and L{\"a}uchli \cite{luscher}, the success of these extrapolations is crucially dependent on the available system sizes. L{\"u}scher and L{\"a}uchli found that extrapolations to determine the spin wave velocity were off when the extrapolation was limited to system sizes of up to 32 spins. Thus while there appears to be a certain robustness of some extrapolation estimates, {\em e.g.}\ for the lowest excitation energies in weakly frustrated systems, other parameters are more directly affected by a small system bias. Furthermore, the ring exchange coupling introduces more excitation degeneracy, which will affect the quality of any extrapolation.

	\begin{figure}
    	\centering
        \includegraphics[width = 7.74cm]{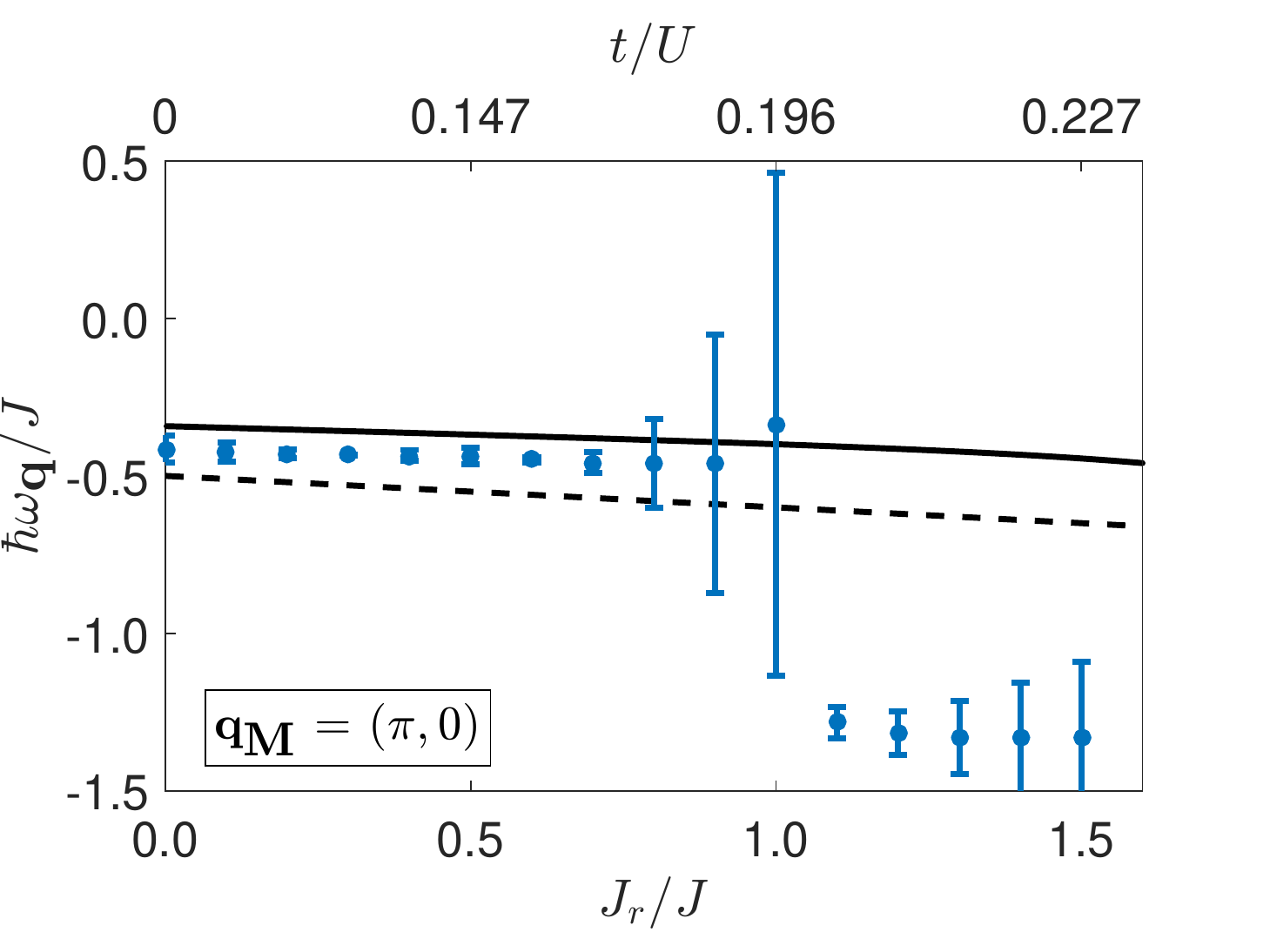}\\ \vspace{0.5cm}
        \includegraphics[width = 7.74cm]{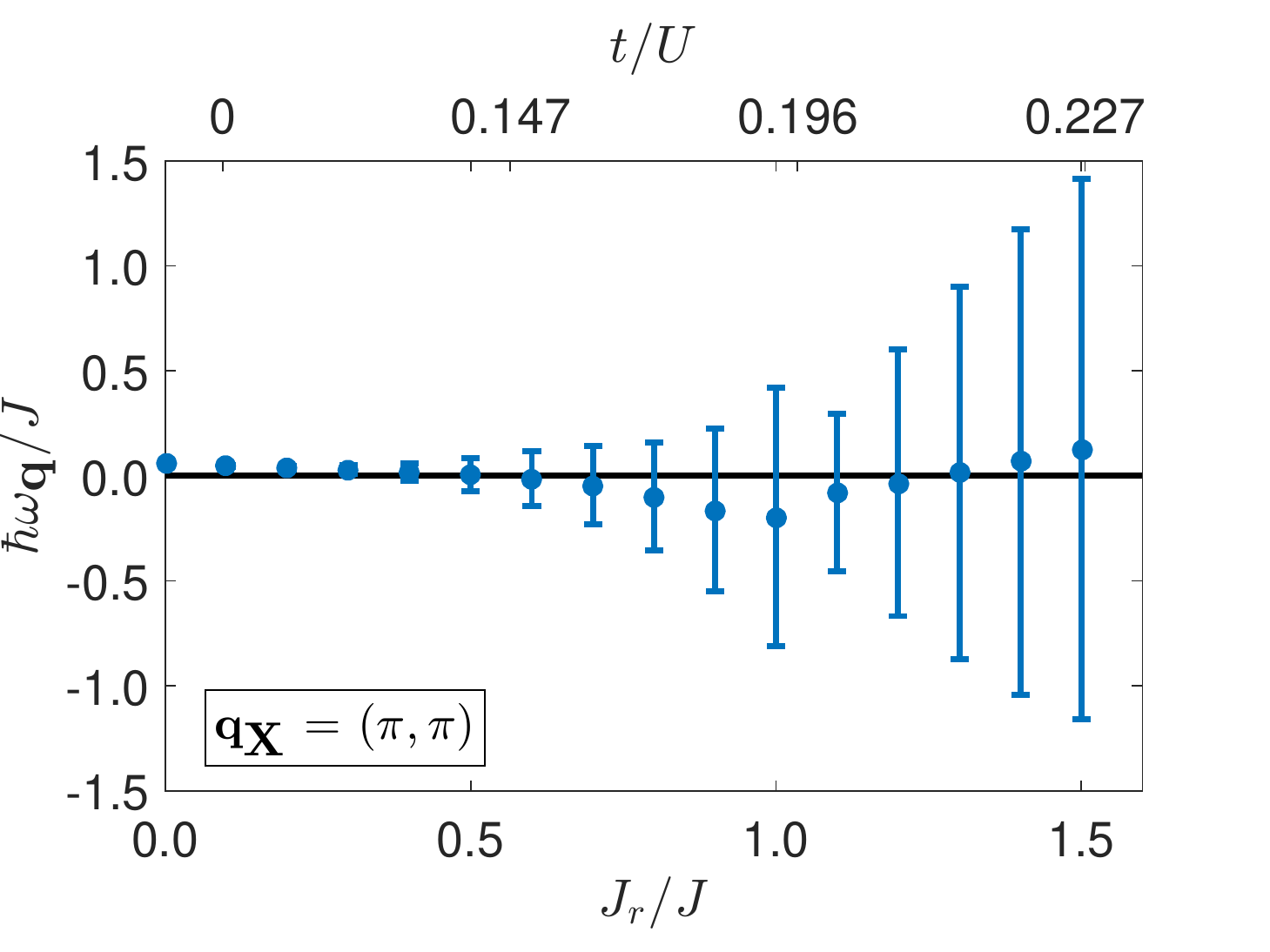}
		\caption{Results of linearly extrapolating exact diagonalization results at the high-symmetry points $\textbf{q}_\textbf{M}$ and $\textbf{q}_\textbf{X}$ to the thermodynamic limit. The dashed lines are the LSW dispersion results from Eq. (\ref{eq:ringdisp}).}
        \label{fig:dispextrap}
    \end{figure}
    
Explorations of different extrapolation schemes of excitation energies at $\textbf{q}_\textbf{M}$ and $\textbf{q}_\textbf{X}$ are shown in Figs.\ \ref{fig:extrappi0} and \ref{fig:extrappipi}.  Due to the limited system sizes, $\textbf{q}_\textbf{M}$ and $\textbf{q}_\textbf{X}$ are some of the few high symmetry wave vectors that are contained in at least 3 system sizes.  The extrapolations have been performed based on the excitations with the most spectral weight, which in the case of $J_r \ll J$ will correspond to the first excitation. Both $\frac{1}{N} \to 0$ and $\frac{1}{\sqrt{N}} \to 0$ have been attempted for various $J_r / J$ couplings. Additionally, extrapolations based on the periodicity along either the horizontal or vertical direction, $p$, have been carried out. The periodicity describes the distance between two equivalent spins in a spin cluster, when periodic boundary conditions are taken into account. Due to the way the differently-sized unit cells are constructed, the periodicity does not increase monotonically with the system size. In the case of system sizes with integer unit cell lengths, such as $N = 16$ with $L = \sqrt{N} = 4$, the periodicity is simply found as $p = L$. However, if $L$ is not an integer, unit cells are constructed as tilted squares, resulting in longer periodicities when the unit cells are tiled to form an infinite lattice. For example, in the $N = 32$ system each equivalent spin can be connected by a (4,4) vector, corresponding to a tilt of $45^\circ$ and a periodicity of 8. On the other hand, equivalent spins in the smaller $N = 26$ system are connected by a (5,1) vector, resulting in a longer periodicity of 26.
    
Figs.\ \ref{fig:extrappi0} and \ref{fig:extrappipi} reveal that the $\frac{1}{N} \to 0$ extrapolations result in parameters with the the lowest statistical errors. The excitation energies derived from these fits are shown in Fig. \ref{fig:dispextrap}. At low $J_r / J$, the extrapolation appear successful and the extrapolated parameters seem to follow the general trend of LSW theory.  However, at around $J_r / J \approx 0.7$, the statistical fitting errors increase massively for all extrapolation methods. At the wave vector $\textbf{q}_\textbf{X}$, the extrapolated energies are even negative for $J_r / J > 0.5$. Keeping in mind that quantum transitions can be size-driven,\cite{bausch17} it is possible that a transition from a mostly  N\'eel ordered ground state to a RVB state happens at different $J_r/J$ couplings for different system sizes. Thus a large statistical error and thus a low-quality linear extrapolation can be an indicator of the spin clusters no longer sharing a common ground state, because some of the spin clusters are more heavily affected by quantum fluctuations than others. 
    
    \begin{figure}
	\centering
	\includegraphics[width =  8.6cm]{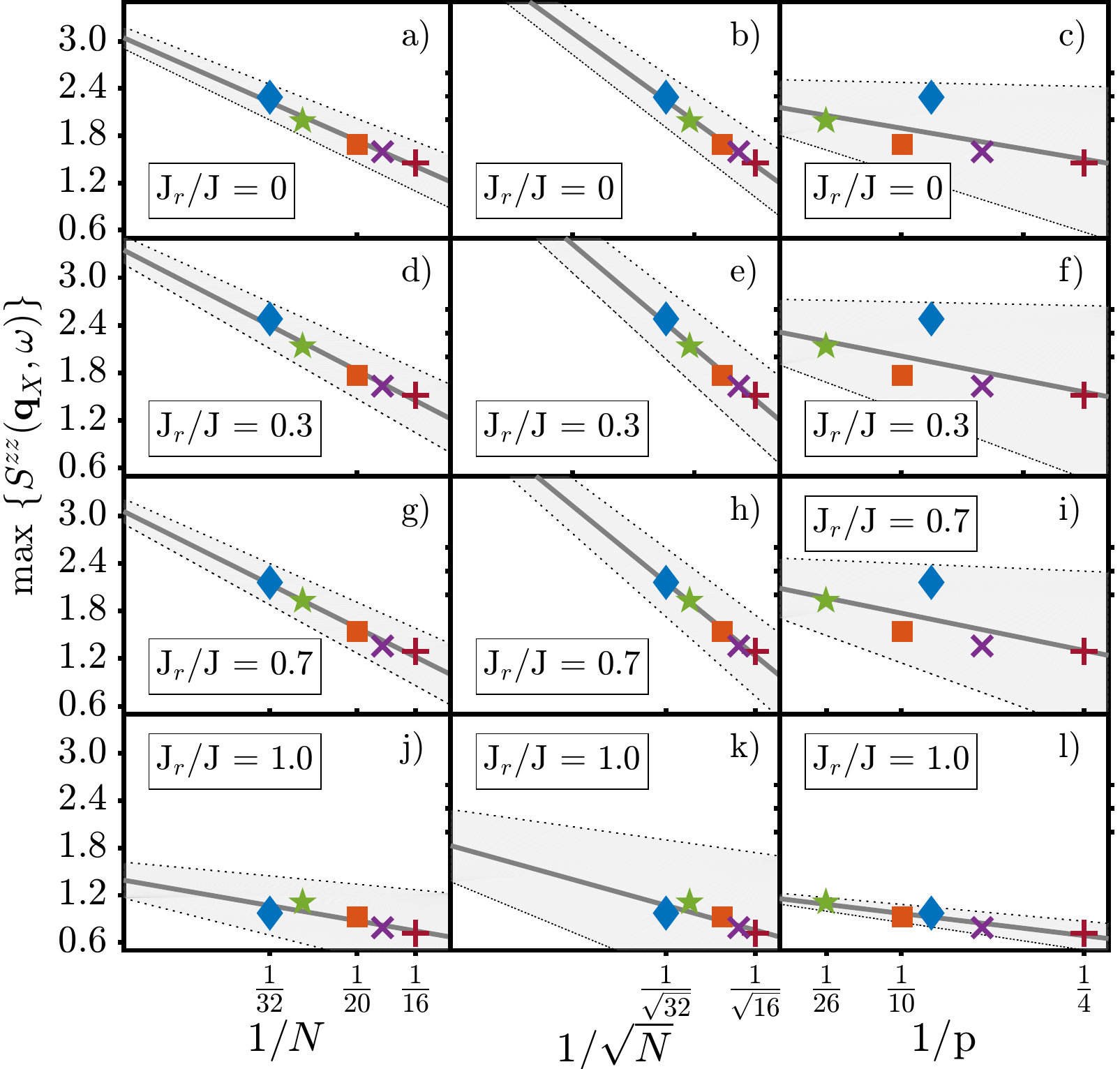}
    \caption{Extrapolations of the largest dynamic structure factors, $S^{zz}(\textbf{q},\hbar \omega)$, at $\textbf{q}_\textbf{X}$ to the thermodynamic limit. }
    \label{fig:extrappipis}
\end{figure}

Fig.\ \ref{fig:extrappipis}	displays similar extrapolations for the largest dynamic structure factors at $\textbf{q}_\textbf{X}$. Overall these extrapolations appear more robust than the energy extrapolation, with system-size dependent ED data being approximately linear even with large $J_r / J$ couplings. All three extrapolation methods indicate that the dynamic structure factor in the thermodynamic limit only varies weakly at $J_r / J \le 0.7$, but at the higher $J_r / J = 1.0$ coupling the value has strongly decreased. This again indicates a destabilization of the magnon modes. The $\frac{1}{N}$ extrapolations mostly results in extrapolated dynamic structure factors with the lowest statistical errors. An exception is the extrapolation shown in Fig. \ref{fig:extrappipis}~l), where the periodicity extrapolation appears to be the most linear. As such, the leading order finite-size effect seems to be more strongly connected to the periodicity and not the total number of spins at high $J_r / J$. Looking back at Fig. \ref{fig:extrappipi}~l), the periodicity extrapolation also results in the thermodynamic excitation energy which is closest to being positive.

    \begin{figure}
	    \centering
	    \includegraphics[width = 7.74cm]{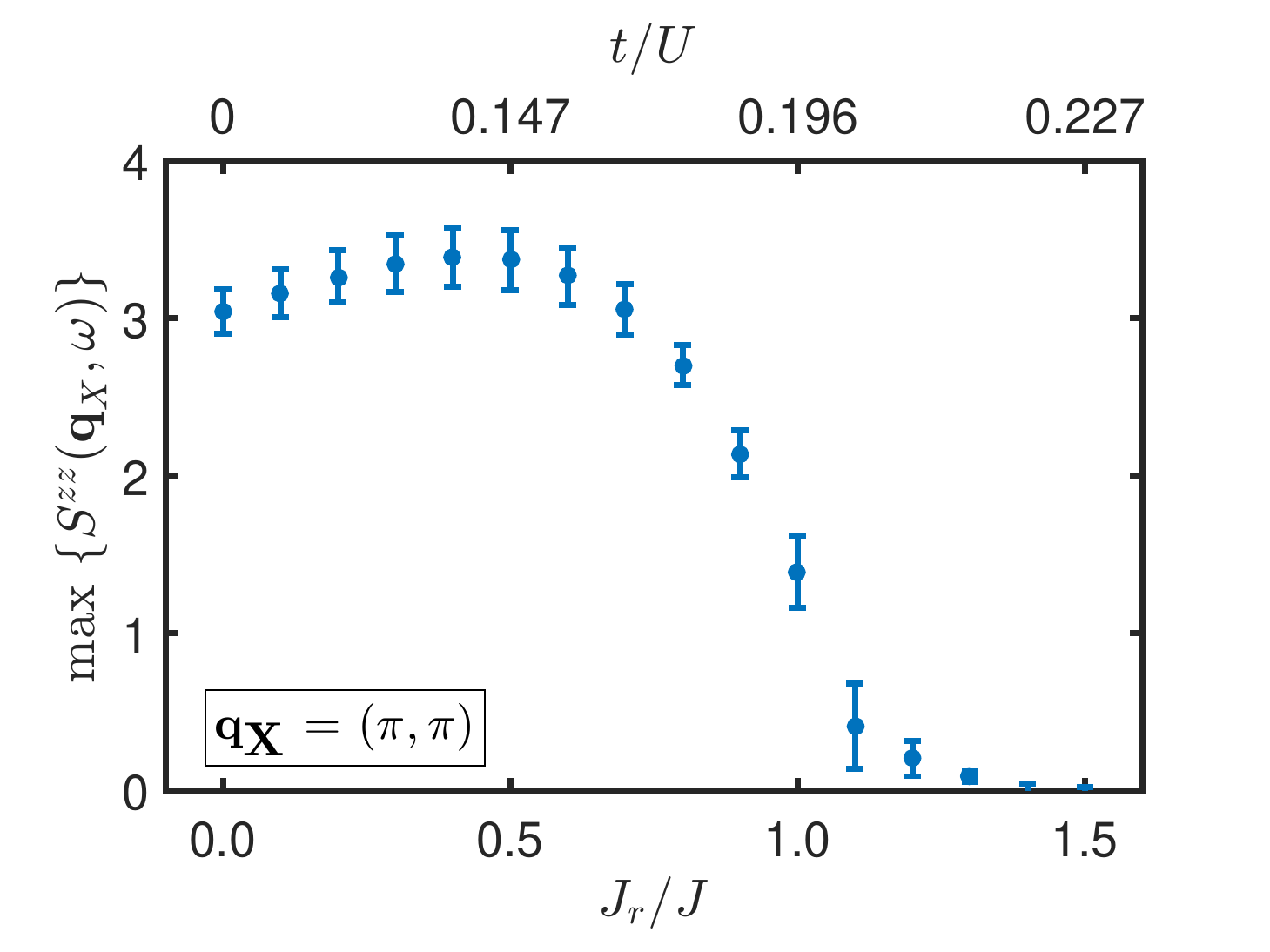}
        \caption{Thermodynamic values of the largest dynamic structure factors, $S^{zz}(\textbf{q}_\textbf{X},\hbar \omega)$, as found with the $\frac{1}{N}$ extrapolations. }
        \label{fig:extrapmaxs}
    \end{figure}
    
The thermodynamic values of the largest dynamic structure factors at $\textbf{q}_\textbf{X}$ are shown as a function of the REM coupling in Fig.~\ref{fig:extrapmaxs}. This parameter would in the thermodynamic limit correspond to the static structure factor $S^{zz}(\textbf{q}_\textbf{X},\omega=0)$. Compared to the extrapolated energy values in Fig.~\ref{fig:dispextrap}, the extrapolated dynamical correlation values appear in general more stable. We observe that the maximum dynamical correlation factor at $\textbf{q}_\textbf{X}$ trends towards zero, strongly reminiscent of a quantum phase transition at $J_r/J \sim 1.1$. We note that this transition point is much lower than the mean field transition of $J_r/J = 2$, the point at which the LSW derived magnetization reaches 0 ($J_r/J \approx 2.7$), or the range in which the LSW gap between the ground state and first excitation energy closes  - at various $\text{q}$-vectors for ($J_r/J \approx 2.0  - 2.5$). 
We ascribe this as the effect of higher order quantum fluctuations lowering the stability of the N\`eel state, as already indicated in Ref.~\onlinecite{majumdar12}.

\section{Conclusion \label{sec:conclusion}}

We have performed a LSW and ED study of the Hubbard-parametrized REM. In contrast to earlier numerical studies of the REM, our ED study has focused on calculating the $S^{zz}(\textbf{q}, \omega)$ values to facilitate an investigation of the dispersion spectrum and the underlying ground state.  The REM LSW dispersion is most uniquely characterized by a higher first excitation energy at $\textbf{q}_\textbf{M}$ than at $\textbf{q}_\textbf{S}$. The same effect is seen in the ED spectrum where the energy difference is found to be greater than in the corresponding LSW spectrum. Furthermore, in systems with strong $J_r / J$ coupling, the first excitation energies of the ED spectrum are found to be lower than the LSW dispersion, indicating a strong quantum renormalization effect. Furthermore the thermodynamic $S^{zz}(\textbf{q}_\textbf{X}, \omega)$ value is seen to decrease at high $J_r / J$ couplings. Another sign of quantum fluctuations is the increased number of states caused by frustration induced by the REM. This is observed directly in the dispersion spectra, where the density of states appears to increase with increased $J_r / J$ coupling.  The formation of a continuum of excited states is in line with RVB studies. 

At low $J_r / J$, extrapolations to the thermodynamic limit have resulted in slightly overestimated first excitation energies at $\textbf{q}_\textbf{M}$ when compared to the LSW dispersion. The extrapolation fits become more unreliable at $J_r / J \ge 0.7$, as observed through the much larger estimated errors. The REM model dispersion is therefore affected by size-dependent quantum effects. 
    
Larger ring exchange couplings ($J_r/J \approx 1.1$) cause the N\`eel state to destabilize due to quantum fluctuations and perform a quantum phase transition to a state with a characteristic wave vector of $\textbf{q}_\textbf{S} = (\pi/2, \pi/2)$. At the quantum critical point, the excitation spectrum is dominated by a number of close lying states and no apparent gap. 

Our results are biased towards smaller system sizes, though signatures of quantum fluctuations, such as an increased number of states, appear to be most pronounced in our largest system size, $N = 32$. 

\begin{acknowledgments}
	We thank the Danish Center for Scientific Computing for providing computational resources. A.T.R. acknowledges support from the Carlsberg Foundation.
\end{acknowledgments}

\bibliography{EDring_paper_manuscript}

\end{document}